\documentclass[9pt,a4paper,fleqn]{article}
\usepackage[a4paper,margin=25.4mm]{geometry}
\usepackage[english]{babel}

\usepackage{cite}
\usepackage{amsmath}
\usepackage{amssymb}
\usepackage{sistyle}
\usepackage{braket}

\usepackage{graphicx}			%% Support for including graphics
\usepackage{wrapfig}			%% Support for figures next to the text, instead of interrupting it
\usepackage[%					%% Change the caption style
	labelfont={sf,bf},			%% Sans serif font, bold style
	textfont={sf},			%% Sans serif font, italics
	font=small,					%% Small font size
	justification=RaggedRight	%% Ragged text
	]{caption}
\usepackage{subcaption}			%% For subfigures
\usepackage{epstopdf}			%% To include .eps figures as pdf
\usepackage{epsfig}
\usepackage{url}

\DeclareMathOperator{\sgn}{sgn}
\newcommand{\ed}{\mathrm d}	
\newcommand{\e}{\mathrm{e}}

\newcommand{\op}[2]{\hat{#1}_{\text{#2}}^{\vphantom{\dag}}}
\newcommand{\hop}[2]{\hat{#1}_{\text{#2}}^{\dag}}
\newcommand{\opp}[3]{{\hat{#1}_{\text{#2}}}^{#3\vphantom{\dag}}}
\begin{document}

\begin{center}
	\hfill\\[1.0cm]
	{\Large\textbf{Mesoscopic Hamiltonian for Josephson travelling-wave parametric amplifiers}}\\[0.2cm]
	%{\Large{A microscopic Hamiltonian of a TWPA}} \\[0.6cm]
	{\large{T.H.A. van der Reep$\left.^{1,*}\right.$\let\thefootnote\relax\footnotetext{$\left.^{*}\right.$e-mail: reep@physics.leidenuniv.nl}}} \\[0.2cm]
	{\emph{Leiden Institute of Physics, Leiden University, Niels Bohrweg $2$, $2333$ CA Leiden, The Netherlands}}\\[0.2cm]
	\today
\end{center}

\begin{abstract}
We present a theory describing parametric amplification in a Josephson junction embedded transmission line. We will focus on the process of four-wave mixing under the assumption of an undepleted pump. However, the approach taken is quite general, such that a different parametric process or the process under different assumptions is easily derived. First the classical theory of the coupled-mode equations as presented by O'Brien \emph{et al.}  [\emph{Phys. Rev. Lett.} $\mathbf{113}:157001$] is shortly reviewed. Then a derivation of the full quantum theory is given using mesoscopic quantisation techniques in terms of discrete mode operators. This results in a Hamiltonian that describes the process of parametric amplification. We show that the coupled-mode equations can be derived from this Hamiltonian in the classical limit and elaborate on the validity of the theory. 
\end{abstract}

\section{Introduction}\label{ChParampsSecIntroduction}
Parametric amplification arises as a result of non-linear optics. In case a non-linear medium is traversed by a (weak) signal and a strong pump, a wave-mixing interaction causes the signal to be amplified. The main advantage of such amplifiers is their low added noise. In comparison: a conventional low-noise microwave amplifier has a noise temperature $T_{\text{n}}$ of several Kelvins, which translates into $k_{\text{B}}T_{\text{n}}/\hbar\omega\approx 10$ photons of added noise for $T_{\text{n}}=\SI{2}{K}$ at a frequency of $\omega/2\pi=\SI{4}{GHz}$ \cite{Bryertonetal2013, LNA, CaltechMic}. This number can be reduced to $1/2$ or even $0$ in a parametric amplifier, depending on its configuration \cite{Caves1982}. This makes parametric amplifiers ideal to amplify signals that are on single-photon level.\\
In the past decade, many microwave parametric amplifiers have been developed to read out quantum bits in quantum information experiments (see e.g. \cite{Guetal2017} for a review). In most of the designs \cite{CastellanosLehnert2007, Bergealetal2010, Rochetal2012, Eichleretal2014, Royetal2015} the amplifier is embedded in a resonator to increase the interaction time of pump and signal, thus to increase the amplifier's gain. Due to such a design these amplifiers, however, are inherently limited in their bandwidth, giving rise to scalability issues now that the number of quantum bits in a single experiment increases. For this reason travelling-wave parametric amplifiers (TWPAs) have been developed \cite{HoEometal2012, Whiteetal2015, Macklinetal2015, Vissersetal2016, Adamhanetal2016, Chaudhurietal2017}. As these are not based on resonance, they do not suffer from the intrinsic bandwidth limitation. However, to achieve a large gain the amplifiers need to be long.\\
Currently, two sources of non-linearity have been considered for TWPAs. Firstly, one can base the amplifier design on the intrinsic non-linear kinetic inductance of superconductors \cite{HoEometal2012, Vissersetal2016, Adamhanetal2016, Chaudhurietal2017}. Secondly, one can embed Josephson junctions in the transmission line, which have a non-linear inductance \cite{Whiteetal2015, Macklinetal2015}. Both versions of the TWPA have been described theoretically using classical coupled-mode equations \cite{HoEometal2012,Whiteetal2015, Yaakobietal2013, OBrienetal2014}. However, a Hamiltonian-description is necessary to describe the TWPA as a quantum device, which is needed for a recently-proposed experiment testing the limits of quantum mechanics by entangling two TWPAs within a single photon interferometer by adding them to each of the interferometer's arms \cite{Reepetal2018}. To interpret the results of such an experiment, it is of utmost importance that one understands precisely how the amplifiers work and what the magnitude of the various coupling constants in the theory are. Moreover, a quantum theory allows to calculate averages, standard deviations and higher-order moments of measurement operators, and takes into account the effects of commutation relations, whereas a classical theory only allows averages to be calculated straightforwardly and the effect of non-commuting operators is neglected. Some authors consider such a Hamiltonian-description impossible due to difficulties of quantum mechanics in describing dispersion \cite{Barralrana2015} (and references therein) -- an important characteristic in TWPAs. However, in case of a TWPA based on Josephson junctions such a description appears to be possible. The Josephson TWPA has already been described using a Hamiltonian based on continuous mode operators \cite{GrimsmoBlais2017}. This description was used to calculate average gain and squeezing effects. In this work we use discrete mode operators for our analysis and use the resulting Hamiltonian to calculate photon number distributions, apart from gain effects.\\
 
We will first put the concept of parametric amplification on solid ground by introducing the necessary terminology. Then, a review is given of O'Brien \emph{et al.} \cite{OBrienetal2014} where the coupled-mode equations were derived, which can be used for predicting the classical response of a TWPA in case the non-linearity in the transmission line is weak. In section \ref{ChParampsSecQuantum} we proceed in deriving the Hamiltonian of the Josephson junction embedded transmission line in the limit of a weak non-linearity, which we apply to the specific case of a non-degenerate parametric amplifier with undepleted degenerate pump in section \ref{ChParampsSecImplementations}. In this section we will also discuss other implementations of the Hamiltonian shortly. Then we will derive the classical coupled-mode equations once more, but now from the quantum Hamiltonian. Thus we show that the classical and quantum theories converge in the classical limit. The chapter is concluded by a discussion of the validity of the theories in terms of the strength of the non-linearity, i.e. to which value of the non-linearity it can be considered weak, in section \ref{ChParampsSSecValidity}.

\section{Terminology}\label{ChParampsSecTerminology}
For parametric amplifiers a specific terminology is used that can be confusing at times. Here an overview of the terminology is presented and it is discussed under which circumstances the terms play a role. These circumstances are fully determined by the Hamiltonian that describes the process.\\
Basically, parametric amplifiers work by the principle of wave mixing. This mixing process occurs due to a non-linear response of the device to a transmitting electromagnetic field and causes energy transfer between the different transmitting modes. Suppose that the non-linearity occurs as a result of a non-linear polarisation of the material,
\begin{equation}
	\mathbf{P}= \left(\chi_{\text{e}}^{\left(1\right)}+\chi_{\text{e}}^{\left(2\right)}\mathbf{E}+\chi_{\text{e}}^{\left(3\right)}\mathbf{E}^2+\dots\right)\mathbf{E},
\end{equation} 
then the Hamiltonian contains a term 
\begin{equation}
	H_{\text{EP}}\propto \mathbf{E}\cdot\mathbf{P} = \chi_{\text{e}}^{\left(1\right)}\mathbf{E}^2+\chi_{\text{e}}^{\left(2\right)}\mathbf{E}^3 +\chi_{\text{e}}^{\left(3\right)}\mathbf{E}^4 +\dots.
\end{equation} 
In case the material has a strong $\chi_{\text{e}}^{\left(2\right)}$-contribution, the $\mathbf{E}^3$-term in the Hamiltonian leads to a three-wave mixing process (3WM) and consequently to a mixing term in the Hamiltonian of the form
\begin{equation}\label{eqH3WM}
	\op{H}{3WM}= \hbar\chi\op{a}{p}\hop{a}{s}\hop{a}{i}\e^{i\left(-\Delta\Omega t +\Delta\phi\right)} + \text{H.c.}.
\end{equation}
This Hamiltonian enables a photon in the pump mode (p) to be scattered into a photon in the signal mode (s) that is to be amplified and some rest energy, which is generally referred to as the idler mode (i). As the Hamiltonian conserves energy, $\omega_{\text{i}}=\omega_{\text{p}}-\omega_{\text{s}}$. Here, $\Delta\Omega$ is a phase-mismatching term resulting from dispersion and modulation in the device, to be discussed in section \ref{ChParampsSecClassical} (equation (\ref{eqAmpEvolClass})) and section \ref{ChParampsSSecParampTermsRev} (equations (\ref{eqDOq1}) and (\ref{eqDOq2})). $\Delta\phi=\phi_{\text{p}}-\phi_{\text{s}}-\phi_{\text{i}}$ is the phase difference between the pump, signal and idler that enter the device.\\
Contrarily, if the material has a dominant $\chi_{\text{e}}^{\left(3\right)}$-contribution, the Hamiltonian contains a term
\begin{equation}\label{eqH4WM}
	\op{H}{4WM} =\hbar\chi\op{a}{p}\op{a}{p'}\hop{a}{s}\hop{a}{i}\e^{i\left(-\Delta\Omega t +\Delta\phi\right)}+ \text{H.c.}
\end{equation}
and a four-wave mixing (4WM) process takes place, where $\Delta\phi=\phi_{\text{p}}+\phi_{\text{p'}}-\phi_{\text{s}}-\phi_{\text{i}}$. In this case two pump photons are scattered into a signal and an idler photon and $\omega_{\text{i}}=\omega_{\text{p}}+\omega_{\text{p'}}-\omega_{\text{s}}$. In case $\op{a}{p}\neq\op{a}{p'}$ the pump is said to be non-degenerate, whereas it is degenerate if $\op{a}{p}=\op{a}{p'}$.\\
Generally, the pump(s) in equations~(\ref{eqH3WM}) and (\ref{eqH4WM}) are treated as classical modes, which are undepleted. This implies that the corresponding operators are replaced by a constant amplitude and can be absorbed in the coupling constant. This results in a contribution to the Hamiltonian that is identical for $3$WM and $4$WM
\begin{equation}\label{eqH34WM_CP}
	\op{H}{3/4WM}=\hbar\tilde{\chi}\hop{a}{s}\hop{a}{i}\e^{i\left(-\Delta\Omega t +\Delta\phi\right)}+ \text{H.c.}
\end{equation}
in which $\tilde{\chi}=\chi\left|A_{\text{p}}\right|$ for $3$WM and $\tilde{\chi}=\chi\left|A_{\text{p}}\right|\left|A_{\text{p'}}\right|$ for $4$WM respectively.

\begin{figure}[htbp]
\centering
	\begin{subfigure}[b]{0.49\textwidth}
	\epsfig{file=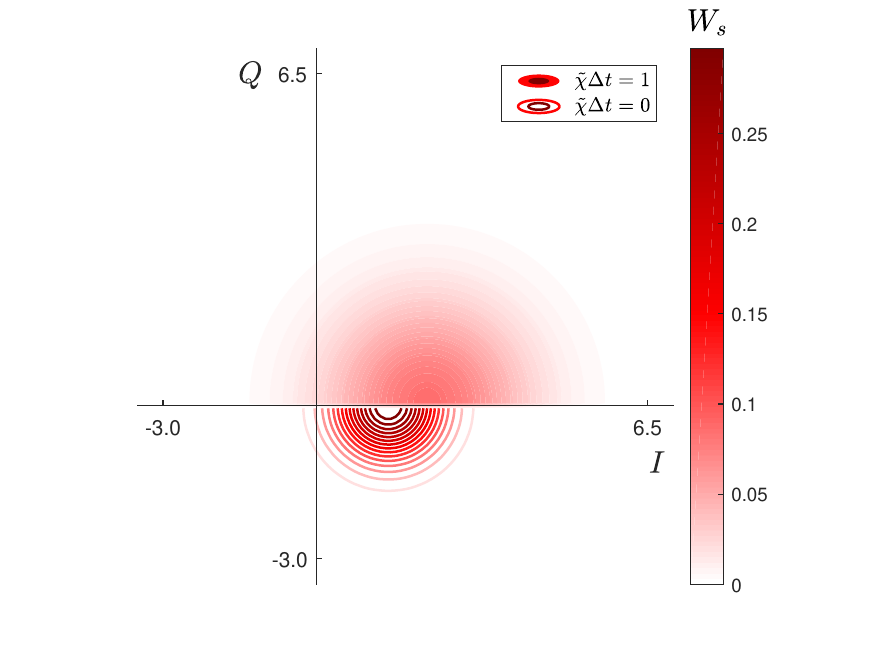, width=\textwidth}
	\caption{}\label{sfigParamp_c_ndeg}
	\end{subfigure}
	\begin{subfigure}[b]{0.49\textwidth}
	\epsfig{file=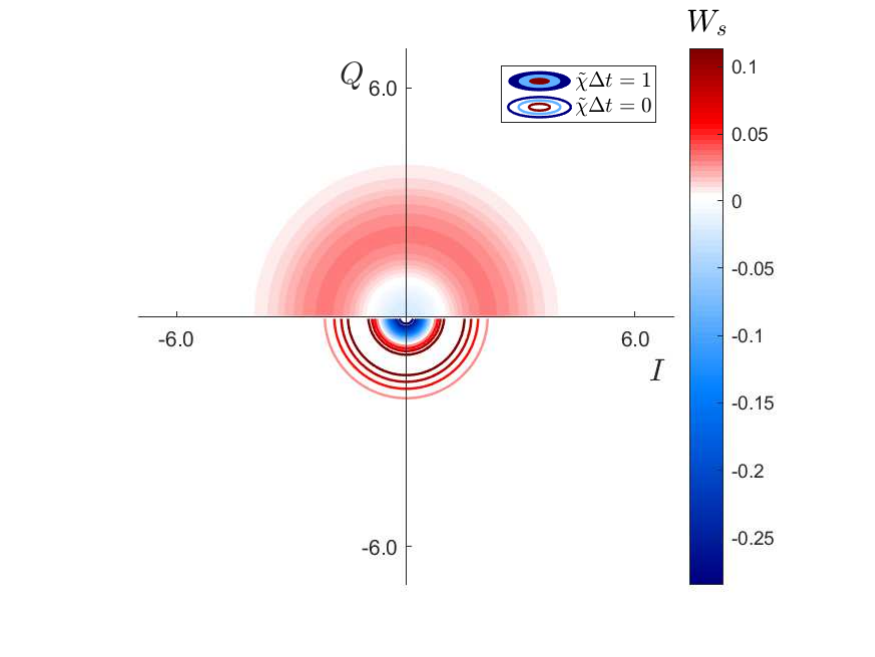, width=\textwidth}
	\caption{}\label{sfigParamp_f_ndeg}
	\end{subfigure}
	\caption{Effect of non-degenerate (or phase-preserving) amplification with an undepleted, degenerate pump on (a) a coherent state ($\alpha=1$) and (b) a single photon number state in the $I,Q$-quadrature plane. The lower half plane depicts (half of the) Wigner function of the unamplified state using contours, whereas the upper half plane shows the Wigner function of the state after amplification with filled contours. The increased width of the latter indicates the increase of noise in the amplified state. The Wigner functions are calculated using \textsc{Qutip} \cite{Qutip}.}\label{figParamp_ndeg}	
\end{figure}

Apart from a distinction in 3WM- and 4WM-devices, parametric amplifiers can be phase-preserving and phase-sensitive. Phase-preserving amplification occurs if the signal and idler are in two distinct modes ($\op{a}{s}\neq\op{a}{i}$ as in equations~(\ref{eqH3WM}) and (\ref{eqH4WM})). For this reason such amplifiers are also referred to as non-degenerate. The amplification is independent of $\Delta\phi$ and a minimum of half a photon of noise per unit bandwidth is added to the signal \cite{Caves1982}. The process is illustrated in figure \ref{figParamp_ndeg}.

\begin{figure}[htbp]
\centering
	\begin{subfigure}[b]{0.49\textwidth}
	\epsfig{file=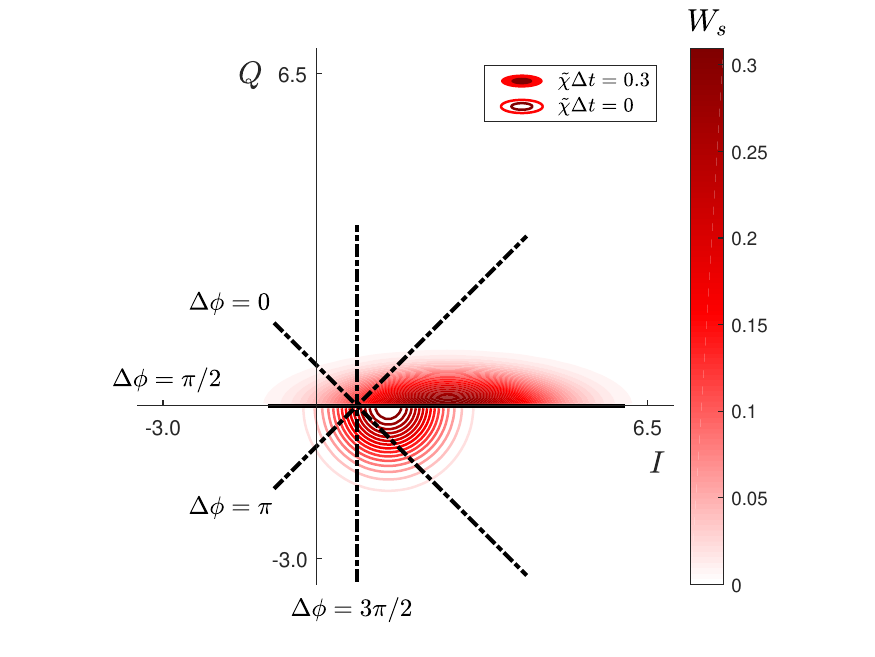, width=\textwidth}
	\caption{}\label{sfigParamp_c_deg}
	\end{subfigure}
	\begin{subfigure}[b]{0.49\textwidth}
	\epsfig{file=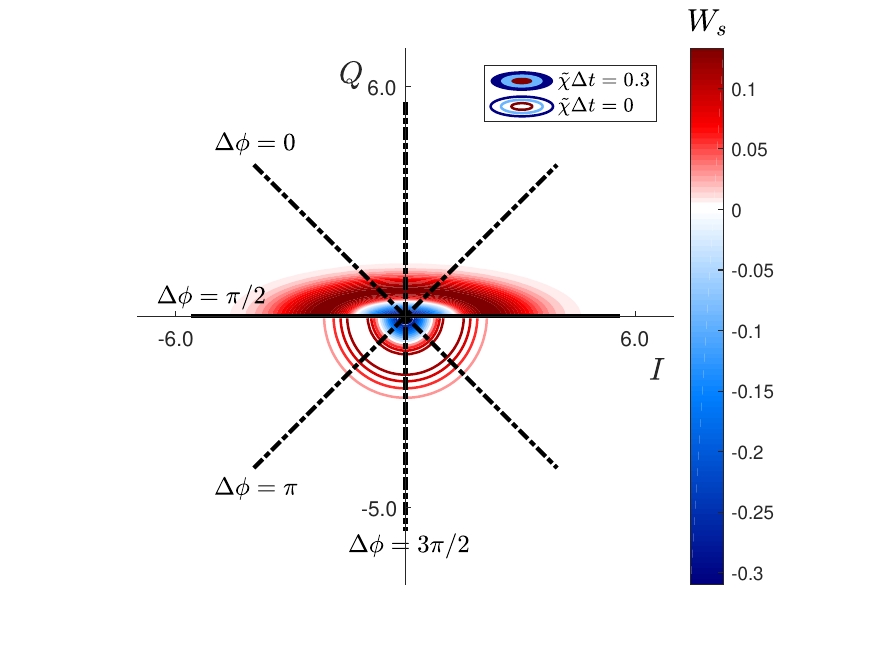, width=\textwidth}
	\caption{}\label{sfigParamp_f_deg}
	\end{subfigure}
	\caption{Effect of degenerate (or phase-sensitive) amplification with an undepleted pump on (a) a coherent state ($\alpha=1$) and (b) a single photon number state in the $I,Q$-quadrature plane. In the lower half plane (half of the) the Wigner functions of the unamplified states are depicted using contours. The upper half plane depicts the Wigner functions of the amplified states for $\Delta\phi=\pi/2$ with filled contours, whereas the long symmetry axis of the Wigner functions for some different $\Delta\phi$s are indictated by dashed lines. The width of one of the quadratures $I$ and $Q$ is amplified, whereas the other is de-amplified, such that the total added noise can be $0$. If the input state is a coherent state, the power gain varies with $\Delta\phi$. The Wigner functions are calculated using \textsc{Qutip} \cite{Qutip}.}\label{figParamp_deg}	
\end{figure}

If signal and idler are in non-distinct modes ($\op{a}{s}=\op{a}{i}$ in equations (\ref{eqH3WM}) and (\ref{eqH4WM})), however, the amplifier is said to be degenerate and works in a phase-sensitive mode. The latter term results from a critical dependence of the amplification process on $\Delta\phi$, which causes one quadrature of the signal to be amplified, whereas the other is de-amplified, see figure \ref{figParamp_deg}. This implies that for input signal states with an explicit phase, such as coherent states, the amplifier's power gain depends on the phase difference between signal and pump. The gain is maximised for $\Delta\phi=\pi/2$ and for $\Delta\phi=3\pi/2$ the gain is less than unity, attenuating the signal. For input states that do not have such an explicit phase, e.g. number states and thermal states, the amplifier power gain is phase-independent. In this process, amplification is possible without adding noise to the signal \cite{Caves1982}.
\\

\begin{figure}[htbp]
	\centering
	\epsfig{file=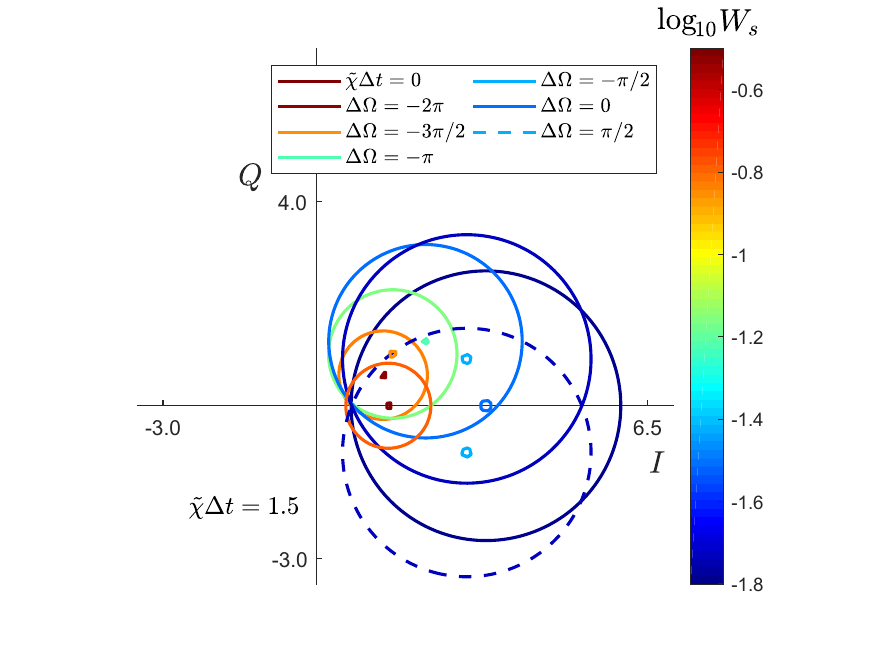, width=0.50\textwidth}
	\caption{Effect of phase mismatch on the phase-preserving amplification of a coherent state for an undepleted pump. Depicted are the full-width-half-maximum and the maximum of the Wigner function of the initial state $\tilde{\chi}\Delta t=0$ and the final states $\tilde{\chi}\Delta t=1.5$ with various amounts of phase mismatch $\Delta\Omega$ ($\Delta t=1$) in the $I,Q$-quadrature plane. The legend refers to the maximum of the Wigner functions. Increasing the phase mismatch reduces the power gain of the amplifier. The Wigner functions are calculated using \textsc{Qutip} \cite{Qutip}.}\label{figParamp_phmismatch}	
\end{figure}

In both 3WM- and 4WM-devices the amplification process is most efficient if the phase mismatch $\Delta\Omega=0$, as is illustrated in figure \ref{figParamp_phmismatch}. A non-zero $\Delta\Omega$ arises from dispersion and modulation effects, which are therefore beneficial to be cancelled.%In microwave devices this can be achieved by e.g. dispersion engineering (cite KIT) or the addition of resonators along the line (cite Obrien and White).

\section{The non-degenerate parametric amplifier with undepleted degenerate pump -- classical theory}\label{ChParampsSecClassical}
\begin{figure}[htbp]
	\centering
	\epsfig{file=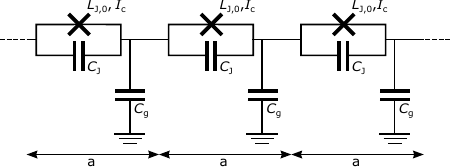, width=0.50\textwidth}
	\caption{Schematic overview of a Josephson junction embedded transmission line. The junctions are modelled as a parallel LC-circuit with a non-linear inductor $L_{\text{J}}$.}\label{figYaakobi_setup}	
\end{figure}

The classical theory for Josephson junction embedded transmission lines is worked out in detail in \cite{Yaakobietal2013,OBrienetal2014}. In \cite{Yaakobietal2013} such a line, as schematically depicted in figure \ref{figYaakobi_setup}, is considered and as a result a non-linear wave equation
\begin{equation}\label{eqNLWaveEq}
	C_{\text{g}}\frac{\partial^2\mathit{\Phi}}{\partial t^2}-\frac{a^2}{L_{\text{J},0}}\frac{\partial^2\mathit{\Phi}}{\partial z^2}-C_{\text{J}}a^2\frac{\partial^4\mathit{\Phi}}{\partial z^2\partial t^2}=
	-\frac{a^4}{2I_{\text{c}}^2 L_{\text{J},0}^3}\frac{\partial^2\mathit{\Phi}}{\partial z^2}\left(\frac{\partial\mathit{\Phi}}{\partial z}\right)^2
\end{equation}
is derived that describes the evolution of the flux $\mathit{\Phi}=\mathit{\Phi}(z,t)$ through the line. Here $C_{\text{g}}$ is the capacitance to ground, $a$ the length of a unit cell of the transmission line, $L_{\text{J},0}$ is the Josephson inductance of the junctions at $0$-flux, $C_{\text{J}}$ is the capacitance of the Josephson junction and $I_{\text{c}}$ its critical current. $L_{\text{J},0}$ and $I_{\text{c}}$ are related by $L_{\text{J},0}=\varphi_0/I_{\text{c}}$ with $\varphi_0=\hbar/2e$ the reduced magnetic flux quantum, see section \ref{ChParampsSecNL-quantum}. In deriving this equation it is assumed that $a\ll\lambda$, the wavelength of the propagating modes, and that the non-linearity provided by the Josephson junctions is weak, such that only the first order non-linear term (right hand side of equation (\ref{eqNLWaveEq})) resulting from the presence of the Josephson junction needs to be taken into account.\\
Starting from this equation, \cite{OBrienetal2014} derives the coupled-mode equations. This is a set of coupled non-linear differential equations that describe the evolution of the considered modes of the flux through the parametric amplifier. For the non-degenerate 4WM parametric amplifier with degenerate pump it is assumed that only three modes of the field play a role. These are generally referred to as the pump, the signal and the idler. The pump is the mode that delivers the energy for the amplification of the small amplitude signal. As a result of energy conservation, an idler mode is created which also has a small amplitude.\\
\cite{OBrienetal2014} suggests a trial solution for equation~(\ref{eqNLWaveEq}) in the form of
\begin{equation}\label{eqTrialSol-NLWave}
	\mathit{\Phi}=\sum_{n=\text{p,s,i}} \text{Re}\left\lbrace A_n\left(z\right) \e^{i\left(k_nz-\omega_nt\right)}\right\rbrace=\frac{1}{2}\sum_n A_{n}\left(z\right)\e^{i\left(k_nz-\omega_nt\right)}+ c.c.
\end{equation}
which describes a superposition of waves that may have varying amplitudes $A_n$ while propagating through the line.\\
Furthermore, the slowly varying amplitude approximation is invoked, i.e., it is assumed that $\left|\ed^2 A_n/\ed z^2\right|\ll \left|k_n \ed A_n/\ed z\right|$, and that the change in amplitude within a wavelength of transmission line is small, $\left|\ed A_n/\ed z\right|\ll \left|k_n A_n\right|$, such that the first order derivatives at the right hand side of equation~(\ref{eqNLWaveEq}) can be neglected. Furthermore, terms of order $|A_{\text{s}}|^2$ and $|A_{\text{i}}|^2$ are neglected, as these are assumed to be small. Then, the amplitudes of the various modes are described by the following differential equations, upon substituting the trial solution into equation~(\ref{eqNLWaveEq}),
\begin{align}
	\frac{\partial A_{\text{p}}}{\partial z} &= i\Xi_{\text{p}}\left|A_{\text{p}}\right|^2A_{\text{p}}+2iX_{\text{p}} A_{\text{p}}^{*}A_{\text{s}}A_{\text{i}}\e^{i\Delta k z} \label{eqParampClassP}\\
	\frac{\partial A_{\text{s(i)}}}{\partial z} &= i\Xi_{\text{s(i)}}\left|A_{\text{p}}\right|^2A_{\text{s(i)}}+iX_{\text{s(i)}} A_{\text{p}}^2A_{\text{i(s)}}^{*}\e^{i\Delta k z} \label{eqParampClassSID}
\end{align}
where $\Delta k =2k_{\text{p}}-k_{\text{s}}-k_{\text{i}}$ and\footnote{Note that this is just the familiar form of $k_n$, $k_n=\omega_n\sqrt{LC}/a$, where $L\mapsto L/\left(1-LC_{\mathbin{\!/\mkern-5mu/\!}L}\omega^2\right)$ as a result of the impedance $Z_{\text{J}}=Z_{L_{\text{J}}}\mathbin{\!/\mkern-5mu/\!}Z_{C_{\text{J}}}$.} 
\begin{equation}\label{eq_knClass}
	k_n=\frac{\omega_n\sqrt{L_{\text{J},0}C_{\text{g}}}}{a\sqrt{1-L_{\text{J},0}C_{\text{J}}\omega_n^2}}.
\end{equation}
The coupling constants $\Xi_n$ and $X_n$ follow to be
\begin{align}
	\Xi_n&=\frac{a^4 k_{\text{p}}^2 k_n^3\left(2-\delta_{\text{p}n}\right)}{16 C_{\text{g}}I_{\text{c}}^2 L_{\text{J},0}^3 \omega_n^2}\label{eqXiClass}	\\
	X_n&=\frac{a^4 k_{\text{p}}^2 k_{\text{s}}k_{\text{i}}\left(k_n-\varepsilon_n \Delta k\right)}{16 C_{\text{g}}I_{\text{c}}^2 L_{\text{J},0}^3 \omega_n^2}\label{eqXClass}
\end{align}
with $\varepsilon_{\text{p}}=1$ and $\varepsilon_{\text{s,i}}=-1$.  As can be noted, the $\Xi_n$s modulate the wave number of the modes in case the pump amplitude is large. $\Xi_{\text{p}}$ is therefore referred to as the self-modulation of the pump, while $\Xi_{\text{s,i}}$ are the cross-modulation terms between the pump and the signal or idler.\\
Under the undepleted pump approximation and assuming $A_{\text{s,i}}\ll A_{\text{p}}$, we can drop the interaction term in equation (\ref{eqParampClassP}) and treat $|A_{\text{p}}|^2$ as a constant. As a result, the equation can be solved analytically as
 \begin{equation}\label{eqPumpDU}
 	A_{\text{p}}=\left|A_{\text{p},0}\right|\e^{i\left(\Xi_{\text{p}}|A_{\text{p},0}|^2 z+\phi_{\text{p}}\right)}.
 \end{equation}
Since we describe 4WM, which is phase preserving, we can assume $\phi_{\text{p}}=0$ with no loss of generality.
 
Substituting this result into equation~(\ref{eqParampClassSID}), it can be rewritten as
\begin{align}
 	\frac{\partial A_{\text{s(i)}}}{\partial z} &= i\Xi_{\text{s(i)}}\left|A_{\text{p},0}\right|^2A_{\text{s(i)}}+iX_{\text{s(i)}}\left|A_{\text{p},0}\right|^2A_{\text{i(s)}}^{*}\e^{i\left(\Delta k+2\Xi_{\text{p}}|A_{\text{p},0}|^2\right) z}\label{eqParampClassSIUD}
\end{align}
Furthermore, switching to a co-rotating frame such that $A_{\text{s(i)}}\mapsto A_{\text{s(i)}}e^{i\Xi_{\text{s(i}}|A_{\text{p},0}|^2 z}$, we can cast the equation in the form
\begin{align}
	\frac{\partial A_{\text{s(i)}}}{\partial z} &= iX_{\text{s(i)}}\left|A_{\text{p},0}\right|^2 A_{\text{i(s)}}^{*}\e^{i\left(\Delta k+\Delta\Xi|A_{\text{p},0}|^2\right) z}\label{eqParampClassSIUDrew1}
\end{align}
where $\Delta\Xi=2\Xi_{\text{p}}-\Xi_{\text{s}}-\Xi_{\text{i}}$. This set of coupled differential equations can be solved analytically as \cite{Armstrongetal1962}
\begin{align}\label{eqAmpEvolClass}
	A_{\text{s(i)}}&=\left[A_{\text{s(i)},0}\left(\cosh g_{\text{z}}z-\frac{i\Delta K}{2g_{\text{z}}}\sinh g_{\text{z}}z\right)+\frac{i X_{\text{s(i)}} \left|A_{\text{p},0}\right|^2}{g_{\text{z}}}A_{\text{i(s)},0}^{*}\sinh g_{\text{z}}z\right]\e^{i\Delta K z/2}
\end{align}
with $\Delta K=\left(\Delta k+\Delta \Xi|A_{\text{p},0}|^2\right)$, which is related to $\Delta\Omega$ in equation (\ref{eqH4WM}) through $k_n+\Xi_n|A_{\text{p},0}|^2\mapsto(k_n+\Xi_n)\omega_n|A_{\text{p},0}|^2/k_n$, see section \ref{ChParampsSSecMarriage}. $g_{\text{z}}=\sqrt{X_{\text{s}}X_{\text{i}}^{*}|A_{\text{p},0}|^4-\left(\Delta K/2\right)^2}$
from which the power gain of the signal for a TWPA of length $l_{\text{T}}$ can be determined as
\begin{align}\label{eqGainClass}
\begin{aligned}
	G_{\text{s}}=\left|\frac{A_{\text{s}}}{A_{\text{s},0}}\right|^2=&\left|\:\cosh g_{\text{z}}l_{\text{T}}-\frac{i\Delta K}{2g_{\text{z}}}\sinh g_{\text{z}}l_{\text{T}}\right|^2+\frac{\left|A_{\text{i},0}\right|^2}{\left|A_{\text{s},0}\right|^2}\left|\frac{X_{\text{s}} \left|A_{\text{p},0}\right|^2}{g_{\text{z}}}\sinh g_{\text{z}}l_{\text{T}}\right|^2+\\
	&+\frac{1}{\left|A_{\text{s},0}\right|^2}\left(-iA_{\text{s},0}A_{\text{i},0}\left(\cosh g_{\text{z}}l_{\text{T}}-\frac{i\Delta K}{2g_{\text{z}}}\sinh g_{\text{z}}l_{\text{T}}\right) \frac{ X_{\text{s}}^{*} \left|A_{\text{p},0}\right|^2}{g_{\text{z}}^{*}}\sinh g_{\text{z}}^{*}l_{\text{T}}+c.c.\right).
\end{aligned}
\end{align}

\subsection{Effect of phase matching}\label{ChParampsSecPhaseMatching}
As noted in section~\ref{ChParampsSecTerminology}, the amplification process is most efficient if $\Delta\Omega=0$. As will become clear from the quantum mechanical treatment of the problem in due course, $\Delta\Omega$ corresponds to $\Delta K$ in equation~(\ref{eqAmpEvolClass}). However, due to dispersion (equation~(\ref{eq_knClass})) and self- and cross-modulation (equation~(\ref{eqXiClass})) this term cannot be $0$ in a transmission line embedded with Josephson junctions. In order to bring it closer to $0$, we need dispersion engineering.\\
\begin{figure}[htbp]
	\centering
	\includegraphics[width=0.3\textwidth]{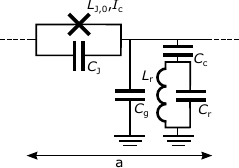}
	\caption{Unit cell of a Josephson junction embedded transmission line with a resonator for achieving phase matching between the pump, signal and idler mode in a TWPA.}\label{figYaakobi_setup_wRes}
\end{figure}
In \cite{OBrienetal2014}, dispersion engineering is achieved by embedding resonators in each unit cell in figure \ref{figYaakobi_setup}, as depicted in figure \ref{figYaakobi_setup_wRes}. If the pump tone is chosen at a frequency close to the resonance frequency of the resonators, the result is that every resonator gives the tone a small phase kick (without diminishing the tone's amplitude too much) and $\Delta K\approx 0$ may be accomplished. The phase kick required per resonator depends on the density of resonators. \cite{OBrienetal2014} puts a resonator in every unit cell, such that the phase kick per resonator only needs to be very small. This implies that only a little of the pump amplitude will be reflected. Contrarily, \cite{Whiteetal2015} puts a resonator after $17$ unit cells each containing $3$ Josephson junctions, such that the required amount of phase shift per resonator is larger, resulting in a larger reflected pump amplitude accordingly.

In case every unit cell contains a resonator, taking its effect on the theory into account is straightforward: we replace the capacitance $C_{\text{g}}$ by an impedance
\begin{equation}
	Z_{C_{\text{eff}},n}=Z_{C_{\text{g}},n}\mathbin{\!/\mkern-5mu/\!}Z_{\text{r},n}=\left(i\omega_n C_{\text{g}}+\frac{i\omega_n C_{\text{c}}\left(1-L_{\text{r}}C_{\text{r}}\omega_n^2\right)}{1-L_{\text{r}}\left(C_{\text{r}}+C_{\text{c}}\right)\omega_n^2}\right)^{-1} 
\end{equation} 
in which $L_{\text{r}}$ and $C_{\text{r}}$ are the inductance and capacitance of the resonator with a resonance at $\omega_{\text{r}}=1/\sqrt{L_{\text{r}}C_{\text{r}}}$. $C_{\text{c}}$ is the coupling capacitance between the resonator and the transmission line. Subsequently we substitute $1/i\omega_n Z_{C_{\text{eff}},n}$ for $C_{\text{g}}$ in the coupling constants $\Xi_n$ and $X_n$ in equations~(\ref{eqXiClass}) and (\ref{eqXClass}), such that
\begin{align}
	\Xi_n^{\text{PM}}&=\frac{ia^4 k_{\text{p}}^2 k_n^3\omega_n Z_{C_{\text{eff}},n}}{16 I_{\text{c}}^2 L_{\text{J},0}^3 \omega_n^2}\left(2-\delta_{\text{p}n}\right)\label{eqXiClassPM}	\\
	X_n^{\text{PM}}&=\frac{ia^4 k_{\text{p}}^2 k_{\text{s}}k_{\text{i}}\omega_n Z_{C_{\text{eff}},n}}{16I_{\text{c}}^2 L_{\text{J},0}^3 \omega_n^2}\left(k_n-\varepsilon_n \Delta k\right).\label{eqXClassPM}
\end{align}

\begin{figure}[htbp]
	\centering
	\epsfig{file=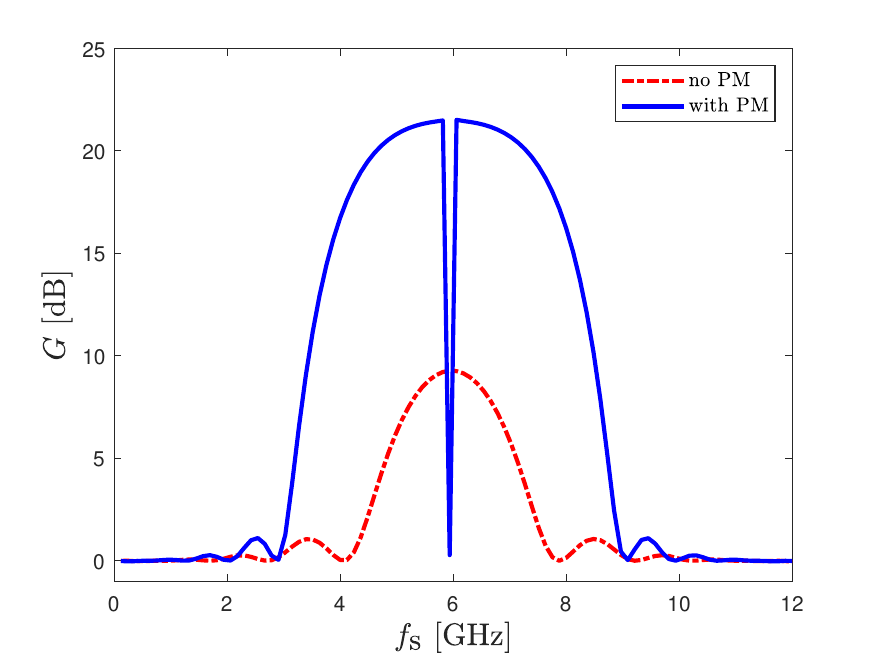, width=0.50\textwidth}
	\caption{Predicted power gain as function of signal frequency without and with phase matching with parameters taken from \cite{OBrienetal2014}. The pump frequency and current are $\SI{5.97}{GHz}$ and $0.5I_{\text{c}}$ and the initial idler current is $0$. The transmission line parameters are $L_{\text{J,$0$}}=\SI{100}{pH}$ ($I_{\text{c}}=\SI{3.29}{\mu A}$), $C_{\text{J}}=\SI{329}{fF}$ and $C_{\text{g}}=\SI{39}{fF}$ and the resonator parameters $C_{\text{c}}=\SI{10}{fF}$, $L_{\text{r}}=\SI{100}{pH}$ and $C_{\text{r}}=\SI{7.036}{pF}$. The calculations have been performed taking into account $2000$ unit cells of $\SI{10}{\mu m}$ length. The dip in the plot for a TWPA with phase matching actually contains two dips on closer inspection. They result from the signal and idler being on resonance with the phase-matching resonators respectively.}\label{figG_class}	
\end{figure}
The effect of phase matching on the performance of the TWPA is depicted in figure \ref{figG_class}. The red dash-dotted curve results from equation (\ref{eqGainClass}) without phase matching with $L_{\text{J,$0$}}=\SI{100}{pH}$ ($I_{\text{c}}=\SI{3.29}{\mu A}$), $C_{\text{J}}=\SI{329}{fF}$ and $C_{\text{g}}=\SI{39}{fF}$ with $2000$ unit cells of $\SI{10}{\mu m}$. The blue continuous curve results from adding resonators to the unit cells and evaluating equation (\ref{eqGainClass}). For the resonators $C_{\text{c}}=\SI{10}{fF}$, $L_{\text{r}}=\SI{100}{pH}$ and $C_{\text{r}}=\SI{7.036}{pF}$. In both calculations the pump current, which is linked to the mode amplitude via the characteristic impedance of the TWPA as $I_{\text{p}}=-A_{\text{p},0}\omega_{\text{p}}/Z_{\text{c}}$, is set to $0.5I_{\text{c}}$ at $\omega_{\text{p}}=2\pi\times\SI{5.97}{GHz}$. The initial idler current is set to $0$.\\ 

In the case that one only adds resonators at specific points in the structure, these can be taken into account by evaluating the amplifier in parts. The resonators divide the structure in sections. Within each section, the evolution of the mode amplitudes follows equation~(\ref{eqAmpEvolClass}) with the coupling constants as given by equations~(\ref{eqXiClass}) and (\ref{eqXClass}). Then, between two sections, one evaluates the transmission coefficient due to the presence of the resonator, updates the mode amplitudes accordingly and uses those amplitudes as input for the next section.

\section{Quantum theory of parametric amplification (4WM)}\label{ChParampsSecQuantum}
In this section we derive the quantum Hamiltonian for the TWPA. In quantum theory the evolution of the state vector, $\ket{\psi}$, describing the system is determined by the Schr\"odinger equation,
\begin{equation}\label{eqSchroedinger}
	i\hbar\frac{\partial \ket{\psi}}{\partial t} = \op{H}{}\ket{\psi}.
\end{equation}
where $\hbar$ is the reduced Planck constant $h/2\pi$. Hence, in order to understand the quantum behaviour of a parametric amplifier, we need to derive its Hamiltonian. In this section the Hamiltonian for a 4WM parametric amplifier, where the non-linearity is provided by Josephson junctions, is derived within Fock space for discrete modes. Although this may sound quite limiting, it should be noted that the same method can be easily applied to three-wave mixing devices or devices with another source of the non-linearity.\\ 
We derive the Hamiltonian in three steps. After covering the concept of energy in transmission lines, a concept which the rest of the derivation relies on, first a dispersionless $LC$-transmission line is quantised. As a second step, dispersion is added to this transmission line by adding an additional capacitance parallel to the inductance. As a final step the inductance is replaced by a Josephson junction.

\subsection{Energy in transmission lines}\label{ssSecEnergyTransmLine}
Typically, non-dissipative transmission lines are quantised as electromagnetic circuits using currents ($I$), fluxes ($\mathit{\Phi}$), voltages ($V$) and charges ($Q$) as quantum fields \cite{VoolDevoret2017}. These give rise to a Hamiltonian via the inductors and capacitors that characterise the line. The energy stored in these elements is given by 
\begin{equation}
	U\left(t\right)=\int_{t_0}^t P \;\ed t' = \int_{t_0}^t VI \;\ed t'
\end{equation}
-- the energy is given by the time-integrated power, $P$, through the element, which equals the product of voltage and current. Now the only task is to calculate the voltage over and current through the element, integrate and sum over all the elements in the circuit. Specifically, for inductors
\begin{equation}\label{eqEnergyL}
	U=
	\begin{cases}
		\int_{I\left(t_0\right)}^{I\left(t\right)} LI \;\ed I' = \dfrac{1}{2}LI^2\\
    	\int_{\mathit{\Phi}\left(t_0\right)}^{\mathit{\Phi}\left(t\right)} \dfrac{1}{L}\mathit{\Phi} \;\ed\! \mathit{\Phi}' = \dfrac{1}{2L}\mathit{\Phi}^2\\ 
    \end{cases}
\end{equation}
using the current-voltage relation for inductors $V=L \partial I/\partial t$ in the first line and Faraday's induction law $V=\partial \mathit{\Phi}/\partial t$ along with $\mathit{\Phi}=LI$ in the second. Note that it is implicitly assumed that the current and flux are $0$ at $t=t_0$. This proves to be a critical assumption of utmost importance as will be shown in section~\ref{ChParampsSecJJCapacitance}. For the energy stored in capacitors, the same form of the energy arises if we interchange current with voltage, flux with charge and inductance with capacitance in equation (\ref{eqEnergyL}).\\

\subsection{Quantisation of a non-dispersive transmission line}
\begin{figure}[htbp]
	\centering
	\includegraphics[width=0.25\textwidth]{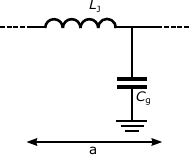}
	\caption{Unit cell of a dispersionless $LC$-transmission line.}\label{figLC_transmissionline}
\end{figure}

Consider once more the transmission line in figure~\ref{figYaakobi_setup}. For the moment we replace the Josephson junction by an ordinary inductor $L_{\text{J}}$ and neglect Josephson capacitance $C_{\text{J}}$, in which case the line is just an ordinary $LC$-transmission line without dispersion, as depicted in figure \ref{figLC_transmissionline}.\\
As suggested by the previous section, we postulate the following mesoscopic Hamiltonian for an electromagnetic (EM) field transmitting through the transmission line
\begin{equation}\label{eqH}
	\op{H}{}=\int_{l_{\text{q}}} \frac{1}{2}\mathcal{L}_{\text{J}}\opp{I}{$L_{\text{J}}$}{2}+\frac{1}{2}\mathcal{C}_{\text{g}}\opp{V}{$C_{\text{g}}$}{2}\; \ed z.
\end{equation}
Here, $\mathcal{L}_{\text{J}}=L_{\text{J}}/a$ is the inductance per unit length and $\mathcal{C}_{\text{g}}=C_{\text{g}}/a$ the capacitance per unit length. $\op{I}{$L_{\text{J}}$}$ is the current through the inductor $L_{\text{J}}$ and $\op{V}{$C_{\text{g}}$}$ is the voltage over the capacitor $C_{\text{g}}$. $l_{\text{q}}$ is the quantisation length \cite{Loudon}.\\
As in the classical theory, we assume sinusoidal waves passing through the line. In this case
\begin{equation}\label{eqV}
	\op{V}{$C_{\text{g}}$}=\sum_n\op{V}{$C_{\text{g}},n$}=\sum_n\sqrt{\frac{\hbar\omega_n}{2\mathcal{C}_{\text{g}}l_{\text{q}}}}\left(\op{a}{$n$}e^{i\left(k_nz-\omega_n t\right)}+H.c.\right),
\end{equation}
as suggested by \cite{VoolDevoret2017}, adapted for discrete mode operators \cite{Loudon}. For waves travelling in positive $z$-direction, the wave number $k_n$ is positive and will be labelled by a positive $n$. Waves travelling in negative $z$-direction have a negative wave number and will therefore be labelled by a negative $n$. In general, $k_{-n}=-k_n$. For frequencies we have $\omega_{-n}=\omega_n$. The characteristic impedance of this line is given by $Z_{\text{c}}=\sqrt{\mathcal{L_{\text{J}}}/\mathcal{C_{\text{g}}}}$ and the phase velocity of the travelling waves equals $v_{\text{ph}}=\omega_n/\left|k_n\right|=1/\sqrt{\mathcal{L_{\text{J}}}\mathcal{C_{\text{g}}}}$ \cite{Pozar}.\\ 
From this voltage we determine the current through the inductor by the telegrapher's equations \cite{Pozar}. Specifically
\begin{equation}
	\frac{\partial V_n}{\partial z}=-\mathcal{L}\frac{\partial I_n}{\partial t}.
\end{equation}
%\begin{align}\label{eqDV}
%	\begin{aligned}
%		\Delta \op{V}{$L_{\text{J}}$}=\sum_n \Delta \op{V}{$n$}&=\sum_n\sqrt{\frac{\hbar\omega_n Z_{\text{c}} v_{\text{ph}}}{2l}}\left(\op{a}{$n$}e^{i\left(k_nz-\omega_n t\right)}\left\lbrace e^{-ik_n a/2}-e^{ik_n a/2}\right\rbrace+H.c.\right)\\
%		&=\sum_n 2\sin\left(k_na/2\right)\sqrt{\frac{\hbar\omega_n Z_{\text{c}} v_{\text{ph}}}{2l}}\left(-i\op{a}{$n$}e^{i\left(k_nz-\omega_n t\right)}+H.c.\right)\\
%		&\approx \sum_n k_na\sqrt{\frac{\hbar\omega_n Z_{\text{c}} v_{\text{ph}}}{2l}}\left(-i\op{a}{$n$}e^{i\left(k_nz-\omega_n t\right)}+H.c.\right).
%	\end{aligned}
%\end{align}
Thus,
\begin{equation}\label{eqI}
	\op{I}{$L_{\text{J}}$}=\sum_n\op{I}{$L_{\text{J}},n$}=\sum_n \sgn(n)\sqrt{\frac{\hbar\omega_n}{2\mathcal{L}_{\text{J}}l_{\text{q}}}}\left(\op{a}{$n$}e^{i\left(k_nz-\omega_n t\right)}+H.c.\right).
\end{equation}\\
Substituting relations~(\ref{eqI}) and~(\ref{eqV}) into equation~(\ref{eqH}) and using that \cite{Loudon}
\begin{align}\label{eqNormalisation}
\begin{aligned}
	\int_{l_{\text{q}}} \e^{i\left(\Delta k_{nm}\right)z} \;\ed z&=
	\begin{cases}
		l_{\text{q}}\;\mathrm{sinc}\left(\Delta k_{nm}l_{\text{q}}/2\right) & \quad \text{if } -l_{\text{q}}/2\leq z \leq l_{\text{q}}/2 \text{ (symmetric bounds)}\\
    	-i\left(\e^{i\Delta k_{nm}l_{\text{q}}}-1\right)/\Delta k_{nm} & \quad \text{if }\qquad\;\, 0\leq z \leq l_{\text{q}} \quad\text{ (asymmetric bounds)}\\ 
    \end{cases}\\
    &\approx l_{\text{q}} \delta_{\Delta k_{nm}}
\end{aligned}
\end{align}
with $\Delta k_{nm}\equiv \pm k_n \pm k_m$. Here, the plus (minus) sign should be chosen if the wave number is associated to an annihilation (creation) operator. The approximation holds if $\Delta k_{nm} l_{\text{q}}\ll 1$, for which
\begin{equation}\label{eqDk-def}
	\delta_{\Delta k_{nm}}=
	\begin{cases}
		1 & \quad \text{if } \Delta k_{nm}l_{\text{q}}=0\\
    	0 & \quad \text{else},\\ 
    \end{cases}
\end{equation}
such that we arrive at
\begin{equation}\label{eqHlin}
	\op{H}{0}=\sum_n \frac{1}{2}\hbar\omega_n\left(\hop{a}{$n$}\op{a}{$n$} +\op{a}{$n$}\hop{a}{$n$}\right) = \sum_n \hbar\omega_n\left(\hop{a}{$n$}\op{a}{$n$} +\frac{1}{2}\right),
\end{equation}
taking into account the commutation relation $\left[\op{a}{$n$},\hop{a}{$m$}\right]=\delta_{nm}$ (See \cite{Loudon} for details). 

\subsection{The influence of the Josephson capacitance: quantisation of a dispersive transmission line}\label{ChParampsSecJJCapacitance}
\begin{figure}[htbp]
	\centering
	\includegraphics[width=0.25\textwidth]{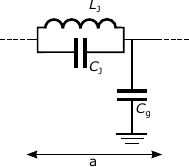}
	\caption{Unit cell of an $LC$-transmission line in which dispersion is added due to the capacitor $C_{\text{J}}$ parallel to the inductor $L_{\text{J}}$.}\label{figLCCj_transmissionline}
\end{figure}

So far we have been neglecting the influence of the parallel capacitor $C_{\text{J}}$ in the transmission line under consideration. Taking this capacitance into account leads to alterations to the theory presented so far, because we now have a capacitor, $C_{\text{J}}$ parallel to the inductor $L_{\text{J}}$, as shown in figure \ref{figLCCj_transmissionline}.
For frequencies $\omega_n<1/\sqrt{L_{\text{J}}C_{\text{J}}}$ this can be taken into account by a frequency-dependent inductance, 
\begin{equation}
	L_{\text{eff}}=\frac{L_{\text{J}}}{1-L_{\text{J}}C_{\text{J}}\omega_n^2}\equiv L_{\text{J}}\Lambda_n,
\end{equation}
and as a result, first
 \begin{equation}\label{eqZcVph_disp}
	Z_{\text{c},n}=\sqrt{\frac{\mathcal{L}_{\text{J}}\Lambda_n}{\mathcal{C}_{\text{g}}}}, \quad v_{\text{ph},n}=\frac{1}{\sqrt{\mathcal{L}_{\text{J}}\Lambda_n\mathcal{C}_{\text{g}}}}=\frac{\omega_n}{\left|k_n\right|},
\end{equation}
implying dispersion is added to the problem, since the phase velocity is now frequency-dependent. Secondly, we have to add an additional capacitive energy to the Hamiltonian.\\

For didactic reasons we now give two derivations of the Hamiltonian in which we take the parallel capacitor into account. In equations (\ref{eqH2}) to (\ref{eqHdispWrong}) we present an erroneous approach, after which the correct manner is presented.\\

Intuitively, the energy contribution of $C_{\text{J}}$ can be added to the Hamiltonian in the same way as the energy stored in $C_{\text{g}}$. That is,
\begin{equation}\label{eqH2}
	\op{H}{}=\int_{l} \frac{1}{2}\mathcal{L}_{\text{J}}\opp{I}{$L_{\text{J}}$}{2}+\frac{1}{2}\mathcal{C_{\text{J}}}\Delta \opp{V}{$C_{\text{J}}$}{2}+\frac{1}{2}\mathcal{C}_{\text{g}}\opp{V}{$C_{\text{g}}$}{2}\; \ed z.
\end{equation}
Realising that $\Delta\op{V}{$C_{\text{J}}$}=\Delta\op{V}{$L_{\text{J}}$}=\Delta\op{V}{$L_{\text{eff}}$}$, it can be calculated by the current-voltage relationship for inductors. The current through $L_{\text{eff}}$ equals the current in equation (\ref{eqI}) with $L_{\text{J}}\mapsto L_{\text{eff}}$. Thus,
\begin{align}\label{eqDV}
	\begin{aligned}
		\Delta \op{V}{$L_{\text{eff}}$}=\sum_n \Delta \op{V}{$L_{\text{eff}},n$}&=L_{\text{J}}\sum_n \Lambda_n\frac{\partial \op{I}{$L_{\text{eff}},n$}}{\partial t}\\
		&= \sum_n k_na\sqrt{\frac{\hbar\omega_n}{2\mathcal{C}_{\text{g}}l_{\text{q}}}}\left(-i\op{a}{$n$}e^{i\left(k_nz-\omega_n t\right)}+H.c.\right).
	\end{aligned}
\end{align}
Using the same relationship, we can calculate $\op{I}{$L_{\text{J}}$}$ as
\begin{equation}\label{eqI_Lj}
	\op{I}{$L_{\text{J}}$}=\frac{1}{L_{\text{J}}}\int \Delta \op{V}{$L_{\text{J}}$}\; \ed t = \sum_n \frac{k_n}{\mathcal{L}_{\text{J}}\omega_n}\sqrt{\frac{\hbar\omega_n}{2\mathcal{C}_{\text{g}}l_{\text{q}}}}\left(\op{a}{$n$}e^{i\left(k_nz-\omega_n t\right)}+H.c.\right)
\end{equation}
and we can use the methods of the last section to find
\begin{equation}\label{eqHdispWrong}
	\op{H}{0}= \sum_n \frac{1}{2}\hbar\omega_n\left(\hop{a}{$n$}\op{a}{$n$} +\frac{1}{2}\right)\left(\Lambda_n+\left(\Lambda_n-1\right)+1\right)=\sum_n \hbar\omega_n\Lambda_n\left(\hop{a}{$n$}\op{a}{$n$} +\frac{1}{2}\right). \qquad \text{(Wrong!)}
\end{equation}
This is an odd result: in the transmission line fed by a mode oscillating at a frequency $\omega_n$ the mode seems to oscillate at $\omega_n\Lambda_n$. Indeed, the result is simply wrong by the exact reason pointed out in section \ref{ssSecEnergyTransmLine}. The voltage in equation (\ref{eqV}) is ``cosine-like'', whereas $\Delta\op{V}{$C_{\text{J}}$}$ is ``sine-like''\footnote{In the sense that the expectation value of the operator on a coherent state $\ket{\alpha}$ with $\alpha \in \mathbb{R}$ scales as either a cosine or a sine.}. This implies that the energy cannot be $0$ in all elements at the same time, as we assumed in equation (\ref{eqEnergyL}). Although the sine-like operator $\Delta\op{V}{$C_{\text{J}}$}$ is unsuitable to be used for the purpose of the derivation of the Hamiltonian, it should be noted that $\Delta\op{V}{$C_{\text{J}}$}$ is a ``valid'' operator in itself and thus it is suitable for calculating expectation values from some quantum state $\ket{\psi}$.\\

To solve this problem, consider once more the energy stored in $C_{\text{J}}$,
\begin{equation}
	U_{C_{\text{J}}}=\int_{t_0}^t V_{C_{\text{J}}}I_{C_{\text{J}}}\; \ed t'=\frac{1}{2}C_{\text{J}}V_{C_{\text{J}}}^2.
\end{equation}
However, we can also cast this energy in terms of the flux, given by Faraday's induction law as
\begin{equation}
	\mathit{\Phi}=\int V \;\ed t. 
\end{equation}
Since the flux is the time integrated voltage, it will be cosine-like, whenever the voltage is sine-like and vice versa. From the definition
\begin{equation}
	U_{C_{\text{J}}}=\int_{t_0}^t V_{C_{\text{J}}}I_{C_{\text{J}}}\;\ed t'=\int_{t_0}^t C_{\text{J}}\frac{\ed\mathit{\Phi}_{C_{\text{J}}}}{\ed t'}\frac{\ed^2\mathit{\Phi}_{C_{\text{J}}}}{\ed t'^2}\;\ed t'=-\frac{\omega^2}{2} C_{\text{J}} \mathit{\Phi}_{C_{\text{J}}}^2,
\end{equation}
using the current-voltage relation for capacitors $I=C\partial V/\partial t$ and that $\partial^2\mathit{\Phi}/\partial t^2=-\omega^2\mathit{\Phi}$. This suggests that a more fruitful approach is to start out with
\begin{equation}\label{eqHdisp}
	\op{H}{}=\frac{1}{2a^2}\int_{l_{\text{q}}} \frac{1}{\mathcal{L}_{\text{J}}}\Delta\!\mathit{\opp{\Phi}{$L_{\text{J}}$}{\mathrm{2}}}+\mathcal{C}_{\text{J}}a^2\Delta\!\mathit{\op{\Phi}{$C_{\text{J}}$}}\frac{\partial^2 \Delta\!\mathit{\op{\Phi}{$C_{\text{J}}$}}}{\partial t^2}+ \frac{1}{\mathcal{C}_{\text{g}}}\opp{Q}{$C_{\text{g}}$}{2}\; \ed z.
\end{equation}
In the above equation we switched to flux and charge variables in all terms for aesthetic reasons. For the first and third term we might use the current and voltage variable just as well. The fluxes can be computed from either equation (\ref{eqDV}) or (\ref{eqI_Lj}), and it follows
\begin{equation}\label{eqPhi_L}
	\Delta\!\mathit{\op{\Phi}{$L_{\text{J}}$}}=\Delta\!\mathit{\op{\Phi}{$C_{\text{J}}$}}=\Delta\!\mathit{\op{\Phi}{$L_{\text{eff}}$}}=\sum_n \frac{k_na}{\omega_n}\sqrt{\frac{\hbar\omega_n}{2\mathcal{C}_{\text{g}}l_{\text{q}}}}\left(\op{a}{$n$}e^{i\left(k_nz-\omega_n t\right)}+H.c.\right).
\end{equation}
Substituting, the Hamiltonian (\ref{eqHdisp}) yields
\begin{equation}\label{eqHdisp_a}
	\op{H}{$0$}=\sum_n \hbar\omega_n\left(\hop{a}{$n$}\op{a}{$n$} +\frac{1}{2}\right)
\end{equation}
as expected.

This result can be generalised for any lossless, linear transmission line. From equation~(\ref{eqHdisp}) we can infer that we can describe the same problem with just two terms in the Hamiltonian. Rewriting equation~(\ref{eqHdisp}) we find
\begin{equation}\label{eqHdisp2}
	\op{H}{}=\frac{1}{2a}\int_{l_{\text{q}}} \left(\frac{1}{L_{\text{J}}}\Delta\!\mathit{\op{\Phi}{$L_{\text{J}}$}}+C_{\text{J}}\frac{\partial \Delta \op{V}{$C_{\text{J}}$}}{\partial t}\right)\Delta\!\mathit{\op{\Phi}{$C_{\text{J}}$}}+ C_{\text{g}}\opp{V}{$C_{\text{g}}$}{2}\; \ed z = \frac{1}{2a}\int_{l_{\text{q}}} \op{I}{$L_{\text{eff}}$}\Delta\!\mathit{\op{\Phi}{$L_{\text{eff}}$}}+C_{\text{g}}\opp{V}{$C_{\text{g}}$}{2}\; \ed z
\end{equation}
as $I_{L_{\text{eff}}}=I_{L_{\text{J}}}+I_{C_{\text{J}}}$. The same argument holds if $C_{\text{g}}$ is replaced by a frequency-dependent effective capacitance $C_{\text{eff}}$, such that we may write 
\begin{equation}\label{eqHdisp3}
	\op{H}{}=\frac{1}{2a}\int_{l_{\text{q}}} \op{I}{$L_{\text{eff}}$}\Delta\!\mathit{\op{\Phi}{$L_{\text{eff}}$}}+\op{V}{$C_{\text{eff}}$}\op{Q}{$C_{\text{eff}}$}\; \ed z
\end{equation}
for any lossless linear transmission line. This yields equation~(\ref{eqHdisp_a}) after substitution of the quantum fields.

\subsection{Adding the non-linearity: quantisation of a non-linear transmission line}\label{ChParampsSecNL-quantum}
\begin{figure}[htbp]
	\centering
	\includegraphics[width=0.25\textwidth]{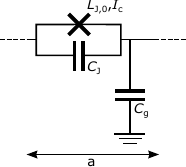}
	\caption{Unit cell of a Josephson junction embedded transmission line in which the Josephson junction is modelled as a non-linear inductor, $L_{\text{J}}$, with a parallel capacitor $C_{\text{J}}$.}\label{figJJCCj_transmissionline}
\end{figure}

As a last step we replace the inductor $L_{\text{J}}$, which we considered as an inductor with a fixed value up to this point, by a Josephson junction. The unit cell for such a transmission line is depicted in figure \ref{figJJCCj_transmissionline}. 
The current through a Josephson junction is \cite{Kittel2004}
\begin{equation}
	I_{\text{J}}=I_{\text{c}}\sin\left(\frac{\Delta\!\mathit{\Phi}_{\text{J}}}{\varphi_0}\right)
\end{equation}
with $\varphi_0=\hbar/2e$ the reduced magnetic flux quantum, $\mathit{\Phi}_0/2\pi$, and $e$ the elementary charge. From the current, we can calculate the Josephson energy in the usual fashion
\begin{equation}
	U_{\text{J}}=\int_{t_0}^t VI \;\ed t'=\int_{t_0}^t \frac{\ed\Delta\!\mathit{\Phi}_{\text{J}}}{\ed t'}I_{\text{c}}\sin\left(\frac{\Delta\!\mathit{\Phi}_{\text{J}}}{\varphi_0}\right) \;\ed t'=I_{\text{c}}\varphi_0\left(1-\cos\left(\frac{\Delta\!\mathit{\Phi}_{\text{J}}}{\varphi_0}\right)\right).
\end{equation}
Substituting this energy for the inductive energy in equation~(\ref{eqHdisp}) yields
\begin{align}\label{eqHnonlin}
\begin{aligned}
	\op{H}{} &=\frac{1}{2a^2}\int_{l_{\text{q}}} 2aI_{\text{c}}\varphi_0\left(1-\cos\left(\frac{\Delta\!\mathit{\op{\Phi}{J}}}{\varphi_0}\right)\right)+\mathcal{C}_{\text{J}}a^2\Delta\!\mathit{\op{\Phi}{J}}\frac{\partial^2 \Delta\!\mathit{\op{\Phi}{J}}}{\partial t^2}+ \frac{1}{\mathcal{C}_{\text{g}}}\opp{Q}{$C_{\text{g}}$}{2}\; \ed z\\
	&=\frac{1}{2a^2}\int_{l_{\text{q}}} \left(\frac{1}{\mathcal{L}_{\text{J},0}}\Delta\!\mathit{\op{\Phi}{J}}-\frac{1}{12\mathcal{L}_{\text{J},0}\varphi_0^2}\Delta\!\mathit{\opp{\Phi}{J}{\mathrm{3}}}+ \mathcal{O}\left({\Delta\!\mathit{\opp{\Phi}{J}{\mathrm{5}}}}\right) +\mathcal{C}_{\text{J}}a^2\frac{\partial^2 \Delta\!\mathit{\op{\Phi}{J}}}{\partial t^2}\right)\Delta\!\mathit{\op{\Phi}{J}}+ \frac{1}{\mathcal{C}_{\text{g}}}\opp{Q}{$C_{\text{g}}$}{2}\; \ed z.
\end{aligned}
\end{align}
in which we have Taylor-expanded the cosine-term and defined the Josephson inductance as $L_{\text{J},0}=\varphi_0/I_{\text{c}}$. From this equation it is clear immediately that the generalised Hamiltonian of equation~(\ref{eqHdisp3}) does not capture the non-linear behaviour.

To address the non-linearity of the transmission line we also calculate the non-linear flux operator derived from equation~(\ref{eqPhi_L}). The dependence of $L_{\text{J}}$ and thus $\Lambda_n$ in the non-linear flux operator on $\Delta\!\mathit{\Phi}_{\text{J}}$ is found from the Josephson current and the flux $\Delta\!\mathit{\Phi}_{\text{J}}=L_{\text{J}}I_{\text{J}}$, 
\begin{equation}\label{eqLnonlin}
	L_{\text{J}}\!\left(\Delta\!\mathit{\Phi}_{\text{J}}\right)=\frac{\varphi_0}{ I_{\text{c}}}\frac{\Delta\!\mathit{\Phi}_{\text{J}}/\varphi_0}{\sin\left(\Delta\!\mathit{\Phi}_{\text{J}}/\varphi_0\right)}\equiv L_{\text{J},0}\frac{\Delta\!\mathit{\Phi}_{\text{J}}/\varphi_0}{\sin\left(\Delta\!\mathit{\Phi}_{\text{J}}/\varphi_0\right)}
\end{equation}
Furthermore, as in the classical theory, we give an explicit time and spatial dependence to the creation and annihilation operators, $\opp{a}{$n$}{(\dag)}\mapsto\opp{a}{$n$}{(\dag)}(z,t)$, in the voltage operator of equation (\ref{eqV}). Invoking the slowly varying amplitude approximation, the time and spatial dependence of these operators is neglected in deriving the other field operators. Hence, from equation~(\ref{eqPhi_L}), we find for the non-linear Josephson junction flux operator
\begin{align}
\begin{aligned}
	\Delta\!\mathit{\op{\Phi}{J}}&=\sum_n a\sqrt{\frac{\hbar\omega_n }{2l_{\text{q}}}\frac{\mathcal{L}_{\text{J},0}\Delta\!\mathit{\op{\Phi}{J}}/\varphi_0}{\sin\left(\Delta\!\mathit{\op{\Phi}{J}}/\varphi_0\right)-\omega_n^2 L_{\text{J}}C_{\text{J}}\Delta\!\mathit{\op{\Phi}{J}}/\varphi_0}}\left(\op{a}{$n$}e^{i\left(k_nz-\omega_n t\right)}+H.c.\right)\\
	&=\sum_n \sqrt{\frac{1}{1-\Lambda_n\sum_{m,l} \Delta\!\mathit{\op{\Phi}{J,$m$}} \Delta\!\mathit{\op{\Phi}{J,$l$}}/6\varphi_0^2+\mathcal{O}\left(\left(\Delta\!\mathit{\op{\Phi}{J}/\varphi_0}\right)^4\right)}} \Delta\!\mathit{\opp{\Phi}{J,$n$}{\mathrm{(0)}}}
\end{aligned}
\end{align} 
with $\Delta\!\mathit{\opp{\Phi}{J,$n$}{\mathrm{(0)}}}$ given by equation~(\ref{eqPhi_L}). In the second line of this equation, we have written explicitly that $\Delta\!\mathit{\opp{\Phi}{J}{\mathrm{2}}}=\sum_{n,m}\Delta\!\mathit{\op{\Phi}{J,$n$}}\Delta\!\mathit{\op{\Phi}{J,$m$}}$. This recurrent relation can be solved iteratively resulting in
\begin{equation}
	\Delta\!\mathit{\op{\Phi}{J}}=\sum_{n}\left(1+\frac{\Lambda_n}{12}\left(\frac{\Delta\!\mathit{\opp{\Phi}{J}{\mathrm{(0)}}}}{\varphi_0}\right)^2+ \mathcal{O}\left(\left(\frac{\Delta\!\mathit{\opp{\Phi}{J}{\mathrm{(0)}}}}{\varphi_0}\right)^4\right)\right)\Delta\!\mathit{\opp{\Phi}{J,$n$}{\mathrm{(0)}}}
\end{equation}
Substitution of this expression in the Hamiltonian of equation~(\ref{eqHnonlin}) yields the Hamiltonian for a $4$WM parametric amplifier where the non-linearity is due to Josephson junctions. Up to first non-linear order (or fourth order in $\Delta\!\mathit{\op{\Phi}{J}}$) we find
\begin{align}\label{eqHtwpaFull}
\begin{aligned}
	\op{H}{TWPA}=& \sum_n \hbar\omega_n\left(\hop{a}{$n$}\op{a}{$n$}+\frac{1}{2}\right)+\\
	&\!+\!\!\!\sum_{n,m,l,k}\!\frac{-i\hbar^2\e^{-i\Delta\omega_{nmlk}t}}{96\mathcal{L}_{\text{J},0}I_{\text{c}}^2 l_{\text{q}}^2\Delta k_{nmlk}}\left(\e^{i\Delta k_{nmlk}l_{\text{q}}}-1\right)\Bigg[\left(\Lambda_m-3L_{\text{J},0}\Lambda_m C_{\text{J}}\omega_k^2\right)\left\lbrace\op{\tilde{a}}{}+\text{H.c.}\right\rbrace_{n\times m\times l\times k}+\\
	&\quad\quad\quad\quad\quad\quad\:
	+2L_{\text{J},0}\Lambda_m C_{\text{J}}\bigg(%\left.2\left\lbrace\op{\tilde{a}}{}+\text{H.c.}\right\rbrace_{n\times m\times l}\left\lbrace-\omega^2\left(\op{\tilde{a}}{}+\text{H.c.}\right)\right\rbrace_k\right.
	\left\lbrace\op{\tilde{a}}{}+\text{H.c.}\right\rbrace_{n}\left\lbrace\omega\left(-i\op{\tilde{a}}{}+\text{H.c.}\right)\right\rbrace_{m\times l}\left\lbrace\op{\tilde{a}}{}+\text{H.c.}\right\rbrace_{k}+\\
	&\quad\quad\quad\quad\quad\quad\quad\quad\quad\!
	\left.+\left\lbrace\op{\tilde{a}}{}+\text{H.c.}\right\rbrace_{n}\left\lbrace\omega\left(-i\op{\tilde{a}}{}+\text{H.c.}\right)\right\rbrace_{m}\left\lbrace\op{\tilde{a}}{}+\text{H.c.}\right\rbrace_{l}\left\lbrace\omega\left(-i\op{\tilde{a}}{}+\text{H.c.}\right)\right\rbrace_{k}\right.\\
	&\quad\quad\quad\quad\quad\quad\quad\quad\quad\!
	\left.+\left\lbrace\op{\tilde{a}}{}+\text{H.c.}\right\rbrace_{n\times m}\left\lbrace\omega\left(-i\op{\tilde{a}}{}+\text{H.c.}\right)\right\rbrace_{l\times k}\right.\bigg)\Bigg].
\end{aligned}
\end{align}

where $\op{\tilde{a}}{$n$}\equiv \sgn{\left(n\right)}\sqrt{\Lambda_n\omega_n}\op{a}{$n$}$ and we have chosen the asymmetric integral bounds of equation (\ref{eqNormalisation}). The subscript $n\cdot m\cdot l\cdot k$ below the braces indicates multiplication, e.g. $\left\lbrace \Lambda\omega\right\rbrace_{n\cdot m}=\Lambda_n\omega_n\Lambda_m\omega_m$. $\Delta k_{nmlk} \equiv \pm k_n\pm k_m \pm k_l \pm k_k$ for the different terms resulting from expansion of the brackets. A plus (minus) sign refers to a corresponding annihilation (creation) operator, e.g. the term $\op{a}{$n$}\hop{a}{$m$}\hop{a}{$l$}\op{a}{$k$}$ corresponds to $\Delta k_{nmlk}= k_n-k_m-k_l+k_k$. Similarly, $\Delta\omega_{nmkl}\equiv\pm\omega_n\pm\omega_m\pm\omega_l\pm\omega_k$.\\

This is the main result of this work. This Hamiltonian describes the full quantum behaviour of Josephson TWPAs up to first non-linear order. However, we will point out two remaining issues and how they may be dealt with. Firstly, this Hamiltonian does not conserve energy a priori. For energy conservation $\Delta\omega_{nmlk}$ must equal $0$, which does not follow necessarily from the equation. At this point we can demand energy conservation by considering only interactions between modes for which $\Delta\omega_{nmlk}=0$. However, one could also reason that $\Delta\omega_{nmlk}\neq 0$ adds to the phase mismatching term $\Delta\Omega$ in equation (\ref{eqH4WM}), which at small magnitudes (compared to $\omega_n$) already greatly reduces the gain of the amplifier. From this argument it follows that $\Delta\omega_{nmlk}\approx 0$. If the latter is the case, and $\Delta\omega_{nmlk}$ is not strictly $0$ this might lead to line broadening of the modes. For now we will assume strict energy conservation and demand $\Delta\omega_{nmlk}=0$.

A second problem in the expression above is the explicit dependence of the mixing term on the quantisation length. This dependence arises both as a consequence of the dispersion in the line as well as an intrinsic dependence of the mixing term that scales as $l_{\text{q}}^{-2}$. Due to dispersion, $\Delta k_{nmlk}$ and $\Delta\omega_{nmlk}$ cannot equal $0$ simultaneously, which introduces an $l_{\text{q}}$-dependence if we demand $\Delta\omega_{nmlk}=0$ for the interacting modes.\\ 
Partly, the two contributions to the quantisation length dependence of the mixing term cancel each other as
\begin{equation}\label{eqNormalisation2}
	\frac{-i\left(\e^{i\Delta k_{nmlk}l_{\text{q}}}-1\right)}{\Delta k_{nmlk}}=l_{\text{q}}\left(1-\frac{i}{2}\Delta k_{nmlk}l_{\text{q}}+\mathcal{O}\left(\left(\Delta k_{nmlk}l_{\text{q}}\right)^2\right)\right).
\end{equation}
The intrinsic dependence of the mixing term on $l_{\text{q}}$ can be further resolved by introducing a classical pump -- see section~\ref{ChParampsSecImplementations}. To deal with the remaining $l_{\text{q}}$-dependence due to dispersion, we can assume that the dispersion effects are small enough such that $\Delta k_{nmlk}\approx 0$, while $\Delta \omega_{nmlk}=0$. We will make this assumption in the following sections. The problem is also resolved considering a transmission line of length $l_{\text{q}}$ for quantisation, of which just a part contains Josephson junctions and using continuous mode quantisation \cite{GrimsmoBlais2017}. 

\section{Implementations}\label{ChParampsSecImplementations}
Using equation~(\ref{eqHtwpaFull}) one can analyse the different implementations of an amplifier. In this section, we will study the non-degenerate amplifier with degenerate pump in detail, the same amplifier implementation that was studied classically in section~(\ref{ChParampsSecClassical}). Treating the pump as a classical mode, we will solve the problem of the explicit appearance of the quantisation length in the mixing coupling constants in section \ref{ChParampsSec_ndegParamp_degPump}. The section ends with a short discussion of other implementations of 4WM amplifiers in section \ref{ChParampsSSecOtherImplementations}.% Then, we will connect the theory to the terminology of parametric amplifiers presented in section \ref{ChParampsSecTerminology} in section \ref{ChParampsSSecParampTermsRev}. Treating all the modes classically it will be shown that we can derive the coupled-mode equations from the quantum theory in section \ref{ChParampsSSecMarriage}. .

\subsection{The non-degenerate parametric amplifier with undepleted degenerate classical pump -- quantum theory}\label{ChParampsSec_ndegParamp_degPump}
As noted in section \ref{ChParampsSecClassical}, for the non-degenerate parametric amplifier with degenerate pump, it is assumed that only three modes, the pump, signal and idler, play a role. Then, from equation~(\ref{eqHtwpaFull}) we can determine the interaction Hamiltonian of the amplifier as
\begin{align}\label{eqHintDquantumFull}
\begin{aligned}
	\op{H}{int}=& \sum_{n,m=\text{p,s,i}} \hbar\xi_{nm}\left(\hop{a}{$n$}\op{a}{$n$}\hop{a}{$m$}\op{a}{$m$}+\frac{c_{m}}{2}\hop{a}{$n$}\op{a}{$n$}+\frac{c_{n}}{2}\hop{a}{$m$}\op{a}{$m$}\right)+
		\hbar\left(\chi\op{a}{p}\op{a}{p}\hop{a}{s}\hop{a}{i} + H.c.\right)+\\
		&+\sum_{n,m=\text{p,s,i}} \frac{c_n c_m\hbar\xi_{nm}}{2\left(2-\delta_{nm}\right)},
\end{aligned}
\end{align}
taking into account the commutation relations explicitly as $c_n\equiv\left[\op{a}{$n$},\hop{a}{$n$}\right]=1$. The first line in this equation determines the dynamics of the amplifier, whereas the second line represents the added zero-point energy. The coupling constants are found to be
\begin{align}
	\xi_{nm} ={}& \frac{\hbar\Lambda_{n}\omega_{n}\Lambda_{m}\omega_{m}}{16I_{\text{c}}^2\mathcal{L}_{\text{J},0}l_{\text{q}}}\left(2-\delta_{nm}\right)\left(1+\Lambda_{\xi_{nm}}\right)\label{eqCCsFullXi}\\
	\chi ={}& \frac{\hbar\Lambda_{\text{p}}\omega_{\text{p}}\sqrt{\Lambda_{\text{s}}\omega_{\text{s}}\Lambda_{\text{i}}\omega_{\text{i}}}}{8I_{\text{c}}^2\mathcal{L}_{\text{J},0}l_{\text{q}}}\bigg(1+\Lambda_{\chi}\bigg)\label{eqCCsFullChi}
\end{align}
where
\begin{align}
	\Lambda_{\xi_{nm}}&\equiv \frac{2}{3}\left(\frac{\Lambda_n}{\Lambda_m}+\frac{\Lambda_m}{\Lambda_n}-2\right)\\
	\Lambda_{\chi}&\equiv \frac{L_{\text{J},0}C_{\text{J}}}{6}
	\Big(\omega_{\text{p}}\omega_{\text{s}}\big(-2\Lambda_{\text{p}}+5\Lambda_{\text{s}}-3\Lambda_{\text{i}}\big)+
	\omega_{\text{p}}\omega_{\text{i}}\big(-2\Lambda_{\text{p}}-3\Lambda_{\text{s}}+5\Lambda_{\text{i}}\big)+
%	&\quad\quad\quad\quad\quad\quad\quad\quad\quad\quad\quad\quad\quad\quad\quad
	\omega_{\text{s}}\omega_{\text{i}}\big(4\Lambda_{\text{p}}-2\Lambda_{\text{s}}-2\Lambda_{\text{i}}\big)\Big)
\end{align}
and $\left.\xi_{nm}\right|_{\Lambda=0}$ implies that $\xi_{nm}$ should be used without the contribution of $\Lambda_{\xi_{nm}}$.\\ 

In equation~(\ref{eqHintDquantumFull}), the $\xi_{n=m}$-term represents the self-modulation and the $\xi_{n\neq m}$-terms represent the cross-modulation. The term in the equation with coupling constant $\chi$ is where the magic happens. This term represents the real amplification process in which two pump photons are scattered into a signal and an idler photon.\\

For a parametric amplifier to work effectively, the device must be driven into its non-linear regime, which is generally achieved by applying a pump current close to the critical current of the device in addition to the much weaker signal current. In this case we can approximate the Hamiltonian to second order in $\opp{a}{s,i}{\left(\dag\right)}$. Moreover, as $\hbar\xi_{nn}\ll\hbar\omega_n$, we can neglect the terms resulting from the commutation relations as well. Hence, to a good approximation,
\begin{equation}\label{eqHDPquantum}
	\op{H}{TWPA}\approx \sum_{n=\text{p,s,i}} \hbar\omega_n\hop{a}{$n$}\op{a}{$n$}+\sum_{n=\text{p,s,i}} \hbar\xi_{\text{p}n}\hop{a}{p}\op{a}{p}\hop{a}{$n$}\op{a}{$n$}+
		\hbar\left(\chi\op{a}{p}\op{a}{p}\hop{a}{s}\hop{a}{i} + H.c.\right),
\end{equation}
where we have neglected the constant zero-point energy, which does not influence the dynamics of the amplifier, and we introduced
\begin{equation}
	\xi_{\text{p}n}= \frac{\hbar\Lambda_{\text{p}}\omega_{\text{p}}\Lambda_{n}\omega_{n}}{16I_{\text{c}}^2\mathcal{L}_{\text{J},0}l_{\text{q}}}\left(4-3\delta_{\text{p}n}\right)\left(1+\Lambda_{\xi_{\text{p}n}}\right).
\end{equation}
Here, the factor $4-3\delta_{nm}$ (instead of $2-\delta_{nm}$) arises from converting the double sum into a single sum. Notably, the coupling constants corresponding to the self- and cross-modulation differ by a numerical factor of $4$. In \cite{GrimsmoBlais2017} this factor was found to be $2$. However, as discussed below, both this work and \cite{GrimsmoBlais2017} find identical operator equations of motion. In equation \ref{eqHDPquantum}, $\chi$ is still given by equation~(\ref{eqCCsFullChi}).\\

Furthermore, we can make the approximation that the pump can be treated as a classical mode and we can replace the corresponding operators with amplitudes. In accordance to the classical treatment of the problem in section~\ref{ChParampsSecClassical}, we will choose the flux, $\op{\Phi}{$C_{\text{g}}$}= \int \op{V}{$C_{\text{g}}$}\; \ed t$, for the amplitude. Upon comparing this expression with its classical analogue (cf. \cite{Loudon}), equation (\ref{eqTrialSol-NLWave}), we find
\begin{equation}\label{eqOptoAmp}
	\op{a}{p}\mapsto -i\sqrt{\frac{\omega_{\text{p}}\mathcal{C}_{\text{g}}l_{\text{q}}}{2\hbar}}A_{\text{p}}.
\end{equation}
Then, for the signal and idler mode, which are still treated quantum mechanically, we find the classical pump Hamiltonian
\begin{equation}\label{eqHDU}
	\opp{H}{TWPA}{\text{(CP)}}\approx \sum_{n=\text{s,i}} \hbar \left(\omega_n+\xi'_{n}\left|A_{\text{p}}\right|^2\right)\hop{a}{$n$}\op{a}{$n$}-
		\hbar\left(\chi' A_{\text{p}}^2\hop{a}{s}\hop{a}{i} + H.c.\right)
\end{equation}
with
\begin{align}
	\xi'_n &= \frac{\Lambda_{\text{p}}\omega_{\text{p}}^2\Lambda_n\omega_{n}}{32I_{\text{c}}^2\mathcal{L}_{\text{J},0}Z_{\text{c},p}v_{\text{ph},p}}\left(4-3\delta_{\text{p}n}\right)\left(1+\Lambda_{\xi_{\text{p}n}}\right)=
	\frac{k_{\text{p}}^2\Lambda_n\omega_{n}}{32I_{\text{c}}^2\mathcal{L}_{\text{J},0}^2}\left(4-3\delta_{\text{p}n}\right)\left(1+\Lambda_{\xi_{\text{p}n}}\right)\label{eqCCsCPxi}\\
	\chi' &= \frac{\Lambda_{\text{p}}\omega_{\text{p}}^2\sqrt{\Lambda_{\text{s}}\omega_{\text{s}}\Lambda_{\text{i}}\omega_{\text{i}}}}{16I_{\text{c}}^2\mathcal{L}_{\text{J},0}Z_{\text{c},p}v_{\text{ph},p}}\left(1+\Lambda_{\chi}\right)=\frac{k_{\text{p}}^2\sqrt{\Lambda_{\text{s}}\omega_{\text{s}}\Lambda_{\text{i}}\omega_{\text{i}}}}{16I_{\text{c}}^2\mathcal{L}_{\text{J},0}^2}\left(1+\Lambda_{\chi}\right)\label{eqCCsCPchi}.
\end{align}
Although $\xi'_{\text{p}}$ does not appear in the classical pump Hamiltonian, it is still defined here for future reference.

Generalising these equations to the case in which resonators are added for dispersion engineering is straightforward. Due to our results in section \ref{ChParampsSecJJCapacitance} this is as easy as making the substitution $C_{\text{g}}\mapsto 1/i\omega_n Z_{C_{\text{eff}}}$ (implicit in $Z_{\text{c},n}$, $v_{\text{ph},n}$, $k_{n}$ and $\Lambda_n$) as discussed in section \ref{ChParampsSecPhaseMatching}.\\

To calculate the gain predicted by a parametric amplifier from the quantum theory, we calculate the Heisenberg equations of motion of the operators. By substituting equation (\ref{eqHDPquantum}) as the Hamiltonian and approximating the pump as a classical mode, this yields
\begin{align}
	\frac{\partial A_{\text{p}}}{\partial t}&=-i\left(\omega_{\text{p}}+2\xi'_{\text{p}}|A_{\text{p}}|^2+c_{\text{p}}\xi_{\text{pp}}\right)A_{\text{p}}+2i\chi^{*}A_{\text{p}}^{*}\op{a}{s}\op{a}{i}\label{eqHeomPtimeQuCP}\\
	\frac{\partial \op{a}{s(i)}}{\partial t}&=-i\left(\omega_{\text{s(i)}}+\xi'_{\text{s(i)}}|A_{\text{p}}|^2\right)\op{a}{s(i)}+i\chi' A_{\text{p}}^2\hop{a}{i(s)},\label{eqHeomSItimeQuCP}
\end{align}
where again we showed the effect of the commutation relations explicitly. However, again we can neglect $\xi_{\text{pp}}$, since $\xi_{\text{pp}}\ll \xi'_{\text{p}}|A_{\text{p}}|^2$. Under the undepleted pump approximation we can neglect the last term in equation~(\ref{eqHeomPtimeQuCP}) as well and solve for the pump amplitude directly, as in the classical case. Hence, in the co-rotating frame introduced in section \ref{ChParampsSecClassical},
\begin{equation}\label{eqHeomSItimeQuRot}
	\frac{\partial \op{a}{s(i)}}{\partial t}=i\chi' \left|A_{\text{p},0}\right|^2\hop{a}{i(s)}\e^{-i\Delta\Omega t}.
\end{equation}
Thus we find, similar to the classical theory
\begin{align}\label{eqAmpEvolQu}
	\op{a}{s(i)}&=\left[\op{a}{s(i),$0$}\left(\cosh g_{\text{t}}t+\frac{i\Delta \Omega}{2g_{\text{t}}}\sinh g_{\text{t}}t\right)+\frac{i \chi' \left|A_{\text{p},0}\right|^2}{g_{\text{t}}}\hop{a}{i(s),$0$}\sinh g_{\text{t}}t\right]\e^{-i\Delta \Omega t/2}
\end{align}
with 
\begin{align}
	\Delta \Omega&= 2\left(\omega_{\text{p}}+2\xi'_{\text{p}}\left|A_{\text{p},0}\right|^2\right)-\left(\omega_{\text{s}}+\xi'_{\text{s}}\left|A_{\text{p},0}\right|^2\right)-\left(\omega_{\text{i}}+\xi'_{\text{i}}\left|A_{\text{p},0}\right|^2\right)=\left(4\xi'_{\text{p}}-\xi'_{\text{s}}-\xi'_{\text{i}}\right)\left|A_{\text{p},0}\right|^2\\ 	
	g_{\text{t}}&=\sqrt{\left|\chi'\right|^2\left|A_{\text{p},0}\right|^4-\left(\Delta \Omega/2\right)^2}
\end{align}
Then, if the state spends a time $t_{\text{T}}$ in the TWPA,
\begin{align}\label{eqGainQu}
\begin{aligned}
	G_{\text{s}}^{\text{q}}=&\frac{\left\langle\hop{a}{s}\op{a}{s}\right\rangle}{\left\langle\hop{a}{s,$0$}\op{a}{s,$0$}\right\rangle}=\\
	=&\left|\:\cosh g_{\text{t}}t_{\text{T}}+\frac{i\Delta \Omega}{2g_{\text{t}}}\sinh g_{\text{t}}t_{\text{T}}\right|^{\:2}+\frac{\left\langle\hop{a}{i,$0$}\op{a}{i,$0$}\right\rangle+1}{\left\langle\hop{a}{s,$0$}\op{a}{s,$0$}\right\rangle}\left|\frac{\chi' \left|A_{\text{p},0}\right|^2}{g_{\text{t}}}\sinh g_{\text{t}}t_{\text{T}}\right|^2+\\
	&+\frac{1}{\left\langle\hop{a}{s,$0$}\op{a}{s,$0$}\right\rangle}\left(-i\left\langle\op{a}{s,$0$}\op{a}{i,$0$}\right\rangle\left(\cosh g_{\text{t}}t_{\text{T}}+\frac{i\Delta \Omega}{2g_{\text{t}}}\sinh g_{\text{t}}t_{\text{T}}\right) \frac{\chi'^{*} \left|A_{\text{p},0}\right|^2}{g_{\text{t}}^{*}}\sinh g_{\text{t}}^{*}t_{\text{T}}+c.c.\right)
\end{aligned}		
\end{align}
in which the term on the second line yields $0$ in case the signal or the idler is initially in a number state.\\

\begin{figure}[htbp]
\centering
	\begin{subfigure}[b]{0.49\textwidth}
	\epsfig{file=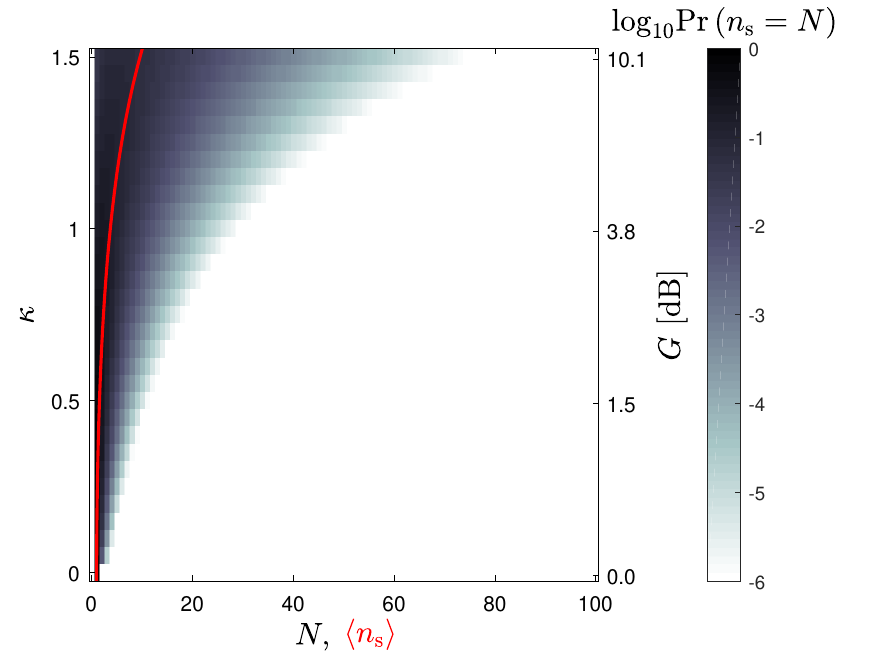, width=\textwidth}
	\caption{}\label{sfigNsdist_fock}
	\end{subfigure}
	\begin{subfigure}[b]{0.49\textwidth}
	\epsfig{file=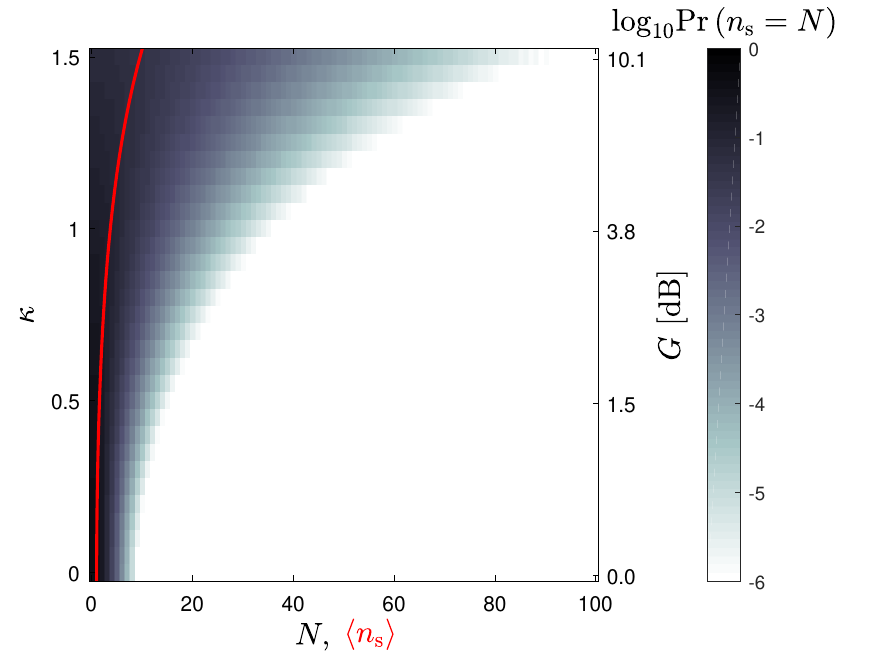, width=\textwidth}
	\caption{}\label{sfigNsdist_coh}
	\end{subfigure}
	\caption{Photon number distribution in the output state of a TWPA for (a) a single photon state and (b) a coherent state $\alpha=1$ as a function of amplification $\kappa=\chi' \left|A_{\text{p}}\right|^2t$ (left axis) or, equivalently, gain $G$ (right axis), assuming $\Delta\Omega=0$. The colourbar, which is cut off at $Pr<10^{-6}$, indicates the probability of finding $N$ photons in the output state. The average number of photons in the output state is indicated in red.}\label{figNsdist}	
\end{figure}

One can also calculate the photon number distribution in the output of a parametric amplifier from the theory in the limit of a classical undepleted pump. To this end we calculate the evolution of the state vector from equation (\ref{eqSchroedinger}) in the interaction picture,
\begin{equation}\label{eqSchroedingerEv}
	\ket{\psi_{\text{I}}\left(t\right)}=\e^{-i\opp{H}{TWPA}{\text{(CP,rot)}} t/\hbar}\ket{\psi_{\text{I}}\left(0\right)}
\end{equation} 
where 
\begin{equation}
	\opp{H}{TWPA}{\text{(CP,rot)}}=-
		\hbar\left(\chi' \left|A_{\text{p}}\right|^2\hop{a}{s}\hop{a}{i}\e^{-i\Delta\Omega t} + H.c.\right)
\end{equation}
is the classical pump Hamiltonian rewritten in the co-rotating frame. Assuming $\Delta\Omega=0$ and $\chi'\in\mathrm{Re}$, we can rewrite the propagator in equation (\ref{eqSchroedingerEv}) using an ordering theorem \cite{BarnettRadmore1997}
\begin{equation}
	\e^{-i\opp{H}{TWPA}{\text{(CP,rot)}} t/\hbar}{|}_{_{\Delta\Omega=0}}=\e^{i\tanh\left(\kappa\right)\hop{a}{s}\hop{a}{i}}\e^{-\ln\left(\cosh\left(\kappa\right)\right)\left(1+\hop{a}{s}\op{a}{s}+\hop{a}{i}\op{a}{i}\right)}\e^{i\tanh\left(\kappa\right)\op{a}{s}\op{a}{i}}
\end{equation}
where the amplification $\kappa\equiv\chi' \left|A_{\text{p}}\right|^2t$.\\
For a single photon input state $\ket{1}_{\text{s}}\ket{0}_{\text{i}}$, we calculate the output state as
\begin{equation}
	\ket{\psi_{\text{I}}\left(t\right)}=\sum_{n=0}^{\infty} \frac{\left(i\tanh\left(\kappa\right)\right)^n}{\cosh^2\left(\kappa\right)}\sqrt{n+1}\ket{n+1}_{\text{s}}\ket{n}_{\text{i}}
\end{equation}
from which we easily compute that the probability of finding $N$ signal photons in the output state equals
\begin{equation}
	\text{Pr}\left(n_{\text{s}}=N\right)=\left|\braket{N|\psi_{\text{I}}\left(t\right)}\right|^2=\frac{\tanh\left(\kappa\right)^{2\left(N-1\right)}}{\cosh^4\left(\kappa\right)}N.
\end{equation}
For a coherent state $\ket{\alpha}_{\text{s}}\ket{0}_{\text{i}}$ we find
\begin{equation}
	\ket{\psi_{\text{I}}\left(t\right)}=e^{-|\alpha|^2/2}\sum_{n,m=0}^{\infty}\frac{\left(i\tanh\left(\kappa\right)\right)^m}{\left(\cosh\left(\kappa\right)\right)^{1+n}}\frac{\alpha^n}{\sqrt{n!}}\sqrt{\binom{n+m}{n}}\ket{n+m}_{\text{s}}\ket{m}_{\text{i}}
\end{equation}
and
%\begin{align}
%\begin{aligned}
%	Pr\left(n_{\text{s}}=N\right)=e^{-|\alpha|^2}\sum_{n,n'=0}^{N} \frac{\left(i\tanh\left(\kappa\right)\right)^{2N-n-n'}}{\left(\cosh\left(\kappa\right)\right)^{2+n+n'}}\frac{\alpha^n\left(\alpha^{*}\right)^{n'}}{\sqrt{n!n'!}}\sqrt{\binom{N}{n}\binom{N}{n'}}.
%\end{aligned}
%\end{align}
\begin{align}
\begin{aligned}
	\text{Pr}\left(n_{\text{s}}=N\right)=e^{-|\alpha|^2}\sum_{n=0}^{N} \frac{\left(\tanh\left(\kappa\right)\right)^{2\left(N-n\right)}}{\left(\cosh\left(\kappa\right)\right)^{2\left(1+n\right)}}\frac{\left|\alpha\right|^{2n}}{n!}\binom{N}{n}.
\end{aligned}
\end{align}
These probabilities are visualised in figure \ref{figNsdist}, in which can be observed how the photon number distribution spreads out as a function of the TWPA gain.

\subsection{Other implementations}\label{ChParampsSSecOtherImplementations}
Although it is most trivial to use the TWPA based on Josephson junctions as a non-degenerate amplifier with degenerate pump, there are other implementation schemes, which will be discussed shortly in this section. As in the last section we assume that only the pump(s) is (are) a source of amplification.  
\begin{itemize}
	\item Non-degenerate pump, signal and idler: Instead of feeding the TWPA with a single pump tone, we can apply two pump tones at different frequencies. In this case we will find, in first order, three (primary) idler tones, one of which will arise as a result of mixing with the two different pump tones $\omega_{\text{i}}=\omega_{\text{p}1}+\omega_{\text{p}2}-\omega_{\text{s}}$, which has been considered theoretically in \cite{OBrienetal2014}. Two other idler tones result from each pump working in a degenerate regime, for which $\omega_{\text{i}1(2)}=2\omega_{\text{p}1(2)}-\omega_{\text{s}}$. However, in turn, each of these idler tones will be the source of (secondary) idler tones at yet different frequencies, such that we end up with a whole spectrum of idlers. A manner to prevent this from happening is to engineer stop bands into the transmission line, such that only specific modes will transmit. However, the generation of many idler tones complicates the analysis for such devices. Still, in general it can be said that such a device will work in the phase-preserving regime.
	\item Non-degenerate pump, signal only: In case we apply two pump tones at different frequencies, we can engineer a quasi phase-sensitive amplifier, if the signal frequency is chosen at $2\omega_{\text{s}}=\omega_{\text{p}1}+\omega_{\text{p}2}$. It will work only in a quasi phase-sensitive regime, because each pump will also cause a primary idler tone to arise from a phase-preserving interaction with the signal. In turn this gives again rise to a whole set of secondary idler tones. If the transmission line is engineered such, that the primary idlers at $\omega_{\text{i}1(2)}=2\omega_{\text{p}1(2)}-\omega_{\text{s}}$ fall into stop bands, the device will work as a real phase-sensitive device.
	\item Pump, signal and idler with DC-current: If we put the TWPA between two bias-Ts we can add a DC-current to the device. In this manner we can use the device as quasi $3$WM, as has been demonstrated in \cite{Vissersetal2016}. Adding the current, we should insert $\mathit{\Phi}_{\text{J}}\mapsto \mathit{\Phi}_{\text{J}}+\mathit{\Phi}_{\text{DC}}$ into the Hamiltonian in equation~(\ref{eqHnonlin}) which yields, among others, a term proportional to $\opp{\mathit{\Phi}}{\text{J}}{3}\mathit{\Phi}_{\text{DC}}$. Continuing the analysis, this yields a term proportional to $\mathit{\Phi}_{\text{DC}}\big(\op{a}{p}\hop{a}{s}\hop{a}{i}+\text{H.c.}\big)$ in the Hamiltonian, which is a $3$WM-term. Of course the full Hamiltonian will contain $4$WM-terms, since the pump acts as a separate source as well. However, choosing the amplitude of the DC-current large with respect to the amplitude of the pump tone, the latter terms can be made small. This implementation of the amplifier is phase-preserving in general, however, it can be used as a phase-sensitive device as well by choosing $\omega_{\text{p}}=2\omega_{\text{s}}$.
\end{itemize}

\section{Terminology -- revisited}\label{ChParampsSSecParampTermsRev}
This work started out with an introduction on the terminology used for parametric amplifiers. After this extensive excursion into 4WM TWPA theory, we have reached the point that we can understand the Hamiltonian in equation~(\ref{eqH4WM}) fully. The only difference between that Hamiltonian and our result in equation~(\ref{eqHDPquantum}) is that the former uses a co-rotating frame, whereas the latter does not. If we cast equation~(\ref{eqHDPquantum}) in a co-rotating frame, we can identify
\begin{align}
	\left.\chi\right|_{\text{eq.(\ref{eqH4WM})}}&=\left.\chi\right|_{\text{eq.(\ref{eqCCsFullChi})}}\\
	\left.\Delta\Omega\right|_{\text{eq.(\ref{eqH4WM})}}&=\left(4\xi_{\text{pp}}-\xi_{\text{ps}}-\xi_{\text{pi}}\right)\hop{a}{p}\op{a}{p}\label{eqDOq1}
\end{align}
or, in case we absorb a classical pump into the coupling constants
\begin{align}
	\left.\tilde{\chi}\right|_{\text{eq.(\ref{eqH34WM_CP})}}&=-\left.\chi'\right|_{\text{eq.(\ref{eqCCsCPchi})}}\left|A_{\text{p}}\right|^2\\
	\left.\Delta\Omega\right|_{\text{eq.(\ref{eqH34WM_CP})}}&=\left(4\xi'_{\text{p}}-\xi'_{\text{s}}-\xi'_{\text{i}}\right)\left|A_{\text{p}}\right|^2.\label{eqDOq2}
\end{align}

\section{Marrying the quantum and classical theories}\label{ChParampsSSecMarriage}
Although the classical theory of the non-linear wave equation and the quantum evolution described by Schr\"odinger's equation seem to be a world apart, in fact the two descriptions can be mapped onto one another. This will be done in this section.\\
The marriage between the two theories runs via the Heisenberg equations of motion of the operators and the connection between the classical mode amplitudes on the one hand and creation and annihilation operators on the other. Starting from the Heisenberg equations of motion with a classical undepleted pump and neglecting small terms (cf. equations (\ref{eqHeomPtimeQuCP}) and (\ref{eqHeomSItimeQuCP}))
\begin{align}
	\frac{\partial A_{\text{p}}}{\partial t}&=-i\left(\omega_{\text{p}}+2\xi'_{\text{p}}|A_{\text{p}}|^2\right)A_{\text{p}}\label{eqHeomPtime}\\
	\frac{\partial \op{a}{s(i)}}{\partial t}&=-i\left(\omega_{\text{s(i)}}+\xi'_{\text{s(i)}}|A_{\text{p}}|^2\right)\op{a}{s(i)}+i\chi' A_{\text{p}}^2\hop{a}{i(s)}.\label{eqHeomSItime}
\end{align}
%we can solve for the pump directly, as in the classical case,
%\begin{equation}
%	A_{\text{p}}=\left|A_{\text{p},0}\right|\e^{-i\left(\omega_{\text{p}}+2\xi_{\text{p}}|A_{\text{p}}|^2\right)t}
%\end{equation}

As the classical coupled-mode equations are defined in space, whereas the Heisenberg equations of motion are equations in time, the first step is to change coordinates from time to space, yielding the spatial Heisenberg equations of motion. From equation~(\ref{eqV}) we can infer that $-\omega_n\partial t=k_n\partial z$ by taking both the derivative to time and to space. Therefore,
\begin{align}
	\frac{\partial A_{\text{p}}}{\partial z}&=i\left(k_{\text{p}}+2\frac{k_{\text{p}}\xi'_{\text{p}}}{\omega_{\text{p}}}|A_{\text{p}}|^2\right)A_{\text{p}}\label{eqHeomPspace}\\
	\frac{\partial \op{a}{s(i)}}{\partial z}&=i\left(k_{\text{s(i)}}+\frac{k_{\text{s(i)}}\xi'_{\text{s(i)}}}{\omega_{\text{s(i)}}}|A_{\text{p}}|^2\right)\op{a}{s(i)}-i\frac{k_{\text{s(i)}}\chi'}{\omega_{\text{s(i)}}} A_{\text{p}}^2\hop{a}{i(s)}.\label{eqHeomSIspace}
\end{align}
%and
%\begin{equation}
%	A_{\text{p}}=\left|A_{\text{p},0}\right|\e^{i\left(k_{\text{p}}+2k_{\text{p}}\xi'_{\text{p}}|A_{\text{p}}|^2/\omega_{\text{p}}\right)z}.
%\end{equation}

As a last step we change the operators back into the classical amplitudes of the modes by virtue of equation~(\ref{eqOptoAmp}). By substitution, we arrive at the classicalised spatial Heisenberg equations of motion
\begin{align}
	\frac{\partial A_{\text{p}}}{\partial z}&=i\left(k_{\text{p}}+\Xi_{\text{p}}^{\text{q}}|A_{\text{p}}|^2\right)A_{\text{p}}\label{eqHeomPspaceClass}\\
	\frac{\partial A_{\text{s(i)}}}{\partial z}&=i\left(k_{\text{s(i)}}+\Xi_{\text{s(i)}}^{\text{q}}|A_{\text{p}}|^2\right)A_{\text{s(i)}}+iX_{\text{s(i)}}^{\text{q}} A_{\text{p}}^2A_{\text{i(s)}}^{*},\label{eqHeomSIspaceClass}
\end{align}
where
\begin{align}
	\Xi_{n}^{\text{q}}&=\frac{k_{n}\xi'_{n}}{\omega_{n}}\left(2-\delta_{\text{p}n}\right) = \frac{a^4 k_{\text{p}}^2 k_n^3\left(2-\delta_{\text{p}n}\right)}{16 C_{\text{g}}I_{\text{c}}^2 L_{\text{J},0}^3 \omega_n^2}\left(1+\Lambda_{\xi_{\text{p}n}}\right)\\
	X_{\text{s(i)}}^{\text{q}}&=\frac{k_{\text{s(i)}}\chi'}{\omega_{\text{s(i)}}}\sqrt{\frac{\omega_{\text{i(s)}}}{\omega_{\text{s(i)}}}}=\frac{a^4 k_{\text{p}}^2 k_{\text{s}}k_{\text{i}}}{16 C_{\text{g}}I_{\text{c}}^2 L_{\text{J},0}^3 \omega_{\text{s(i)}}^2}\left(1+\Lambda_{\chi}\right).
\end{align}

\begin{figure}[htbp]
	\centering
	\epsfig{file=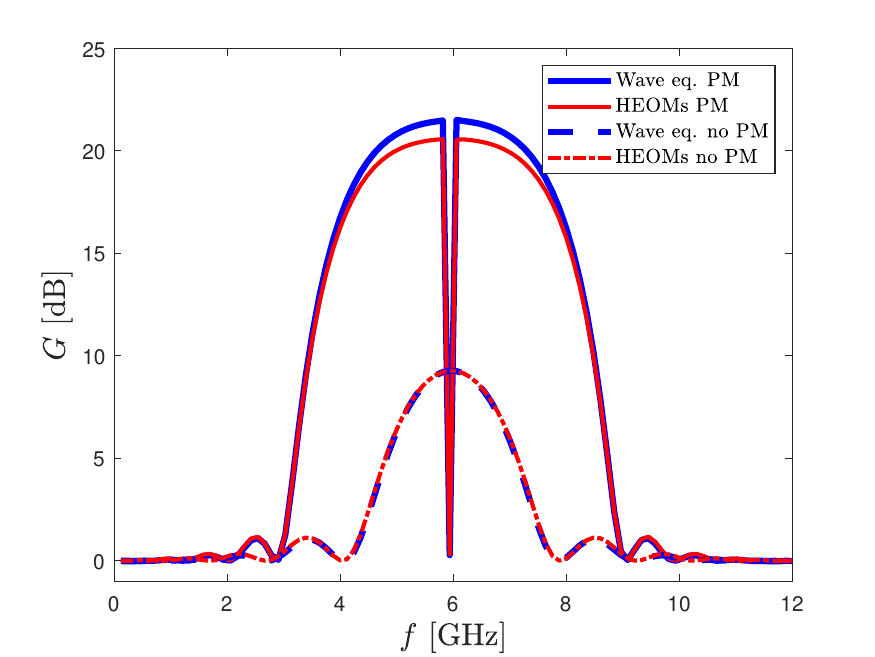, width=0.50\textwidth}
	\caption{Comparison of the predicted power gain from the coupled-mode equations derived from the classical non-linear wave equation and the classicalised spatial Heisenberg equations of motion from the quantum theory. The comparison is made for the case with and without phase matching using the same parameters as in figure \ref{figG_class}. The difference in gain with phase matching is due to the $\Delta k/k_n$-term present in the classical coupled-mode equation coupling constant $X_n$, but absent in the classicalised spatial Heisenberg equation of motion coupling constant $X_n^{\text{q}}$.}\label{figG_CMEvsHEOM}	
\end{figure}

%\begin{align}
%	\frac{\ed A_{\text{(s,i)}}}{\ed z}&=i\left(k_{\text{(s,i)}}+\frac{k_{\text{(s,i)}}\xi'_{\text{(s,i)}}}{\omega_{\text{(s,i)}}}|A_{\text{p}}|^2\right)A_{\text{(s,i)}}+i\frac{k_{\text{(s,i)}}\chi'}{\omega_{\text{(s,i)}}}\sqrt{\frac{\omega_{\text{(i,s)}}}{\omega_{\text{(s,i)}}}} A_{\text{p}}^2A_{\text{(i,s)}}^{*},\label{eqHeomSIspaceClass}
%\end{align}
This set of equations is identical to equations (\ref{eqParampClassP}) and (\ref{eqParampClassSID}) after mapping $A_n\mapsto A_n\e^{ik_n z}$ and removing small terms up to some details: $\Xi_n^{\text{q}}$ contains a factor $\left(1+\Lambda_{\xi_{\text{p}n}}\right)$ which $\Xi_{n}$ does not. Additionally, the factor $(1-\Delta k/k_{\text{s(i)}})$ in $X_{\text{s(i)}}$ has been replaced by $\left(1+\Lambda_{\chi}\right)$ in $X_{\text{s(i)}}^{\text{q}}$. The contributions $\Lambda_{\xi_{\text{p}n}}$ and $\Lambda_{\chi}$ in these factors results from taking into account the non-linear Josephson inductance in $\Lambda_n$ explicitly. The contribution scaling as $\Delta k/k_n$ in $X_n$, however, cannot arise from the quantum theory, since it would need to arise from a coupling constant $\chi$ in the Hamiltonian which is somehow different for the signal and idler mode. Such a difference is not permitted by the quantum theory. However, if we would not have neglected the contribution to $\chi$, $\chi'$ and thus $X_n^{\text{q}}$ due to dispersion, those coupling constants would have been multiplied, up to first order in $\Delta k$, by $\left(1-i\Delta k l_{\text{q}}/2\right)$, see equation (\ref{eqNormalisation2}). This term resembles the factor $(1-\Delta k/k_{\text{s(i)}})$ from the classical theory, although it depends on the unphysical quantisation length. Still, the prediction of gain of both the classical coupled-mode equations and the classicalised spatial Heisenberg equations of motion agree well. As can be observed in figure \ref{figG_CMEvsHEOM}, in case we do not add dispersion engineering there is hardly a difference in predicted gain, whereas only the maximum gain differs in both approaches in case we add dispersion engineering. This is solely due to the factor $\Delta k/k_{\text{s(i)}}$ in $X_{\text{s(i)}}$.

\section{Validity}\label{ChParampsSSecValidity}
In the presented theory we made several assumptions. Firstly, we only took the first non-linear contribution of the Josephson energy into account and secondly it was assumed that the pump can be treated as a classical mode, which is undepleted. In this section, the implications of these assumptions will be presented.\\
 
The theory presented above is derived from a first-order Taylor expansion of the Josephson energy. This implies that at a certain magnitude of the flux through the junction, the theory becomes invalid as higher order terms need to be taken into account. To estimate this flux, we inspect once more the Josephson energy,
\begin{equation}
	U_{\text{J}}=I_{\text{c}}\varphi_0\left(1-\cos\left(\frac{\Delta\!\mathit{\Phi}_{\text{J}}}{\varphi_0}\right)\right)=I_{\text{c}}\varphi_0\sum_{n=1}^{\infty} \frac{\left(-1\right)^{n-1}}{\left(2n\right)!}\left(\frac{\Delta\!\mathit{\Phi}_{\text{J}}}{\varphi_0}\right)^{2n}.
\end{equation}
Thus, we find that the second-order ($n=3$) non-linear effects are approximately $4!(\Delta\!\mathit{\Phi}_{\text{J,p}}/\varphi_0)^2/6!$ smaller than the lowest-order non-linear terms. Hence, if we require that the contribution to the energy of the second-order terms is less than $5\%$ of the energy contribution of the first-order terms, we find that our theory breaks down at $\Delta\!\mathit{\Phi}_{\text{J,p}}/\varphi_0\approx 1.2$ (or $I_{\text{p}}/I_{\text{c}}\approx 0.78$). The dominant second-order amplification term causes two secondary idler modes to appear at $\omega_{\text{(i'),\{i''\}}}=4\omega_{\text{p}}-\omega_{\text{(s),\{i\}}}$ implying the general form of the Hamiltonian in equations~(\ref{eqH4WM}) and~(\ref{eqHDU}) becomes invalid. Moreover, the second-order terms cause additional modulation effects. It is only in the third-order non-linear terms that $(\hop{a}{s}\hop{a}{i})^2$-contributions start to play a role. Furthermore, additional secondary idlers are generated and the modulation effects are further increased. These terms have a maximal energy contribution of approximately $4!(\Delta\!\mathit{\Phi}_{\text{J,p}}/\varphi_0)^4/8!\approx 4\times 10^{-3}$ at the critical flux. Therefore they can be neglected for all practical purposes.\\ 

The undepleted pump approximation breaks down, if the flux of signal and idler photons in the amplifier becomes close to the flux of pump photons. Typically, this happens when $I_{\text{s,}0}\geq I_{\text{p,}0}/10$ \cite{Vissersetal2016,OBrienetal2014}, in which the case the full coupled-mode equations of equations (\ref{eqParampClassP}) and (\ref{eqParampClassSID}) need to be considered to calculate the output amplitudes. Alternatively, for the quantum case the full Hamiltonian of equation (\ref{eqHintDquantumFull}) is to be considered to evaluate the evolution of the quantum state.

\section{Conclusions}\label{ChParampsSecConclusions}
After an introduction to the relevant terminology and the classical theory of the coupled-mode equations of Josephson travelling-wave parametric amplifiers, we derived the mesoscopic quantum Hamiltonian up to first non-linear order describing the process using discrete mode operators. We found that such a description is possible, even when taking into account dispersion effects in the transmission line and showed that the classical coupled-mode equations can be derived from this Hamiltonian. Such a Hamiltonian-description of TWPAs is necessary for treating the amplifier as a quantum device. From our theory we obtain expressions for the coupling constants and can straightforwardly calculate averages, standard deviations and higher-order moments of measurement operators taking into account the commutation relations between operators explicitly. In contrast, from a classical theory only average values follow and the effects of non-commuting operators are neglected.\\
In the derivation, however, there are a few remaining issues. Firstly, it was found that in the non-linear terms of the Hamiltonian, energy and momentum conservation could not be fulfilled simultaneously. Furthermore, the non physical quantisation length is inherent to the theory. The latter can be solved solely under the approximation of a classical undepleted pump. However, the matter can also be resolved in case one derives the Hamiltonian using continuous modes and a transmission line, of which only a part contains the non-linearity. For the concurrent conservation of energy and momentum, we have not been able to find a satisfactory solution.\\
We found that our Hamiltonian, and therefore the coupled-mode equations, are valid to pump currents up to approximately $0.78I_{\text{c}}$. For larger pump currents more non-linear orders have to be taken into account, for which the same recipe can be followed as shown in this paper. The same recipe can also be followed to derive the Hamiltonian for TWPAs that have another source of non-linear behaviour, such as kinetic inductance.\\ 
To make the theories more applicable to experimental realisations of TWPAs, we suggest that the theories can be expanded by taking into account losses as well as reflections within the device and reflections at the boundaries of the device at which it is coupled to its environment.

\subsection*{Acknowledgements}
We would like to thank G. Nienhuis and M.J.A. de Dood for valuable discussions and suggestions. We are grateful to T.H. Oosterkamp for carefully proofreading the manuscript. We also express our gratitude to the Frontiers of Nanoscience programme, supported by the Netherlands Organisation for Scientific Research (NWO/OCW), for financial support.\\

\bibliographystyle{unsrt}
\bibliography{ParampTheory_PRA_V3.bbl} 
\end{document}